\documentclass[12pt]{iopart}
\usepackage{graphicx}
\usepackage{cite}

\eqnobysec

\begin{document}

\title{A model of compact polymers on a family of three-dimensional fractal lattices}

\author{Du\v sanka Leki\' c$^1$ and Sun\v cica Elezovi\' c-Had\v zi\' c$^2$}

\address{$^1$ University of Banja Luka, Faculty of Science, Department of Physics, M.~Stojanovi\' ca 2, Banja Luka,
Bosnia and Herzegovina} \address{$^2$ Faculty of Physics, University of Belgrade, P.O.Box 44, 11001 Belgrade, Serbia}

\eads{\mailto{dusamar@netscape.net}, \mailto{suki@ff.bg.ac.rs}}

\begin{abstract} We study Hamiltonian walks (HWs) on the family of three--dimensional modified Sierpinski gasket  fractals, as a model for compact polymers in nonhomogeneous media in three dimensions. Each member of this fractal family is labeled with an integer $b\geq 2$. We apply an exact recursive method which allows for explicit enumeration of extremely long Hamiltonian walks of different types: closed and open, with end-points anywhere in the lattice, or with one or both ends fixed at the corner sites, as well as some Hamiltonian conformations consisting of two or three strands. Analyzing large sets of data obtained for $b=2,3$ and 4,  we find that numbers $Z_N$ of Hamiltonian walks, on fractal lattice with $N$ sites, for $N\gg 1$ behave as $Z_N\sim \omega^N \mu^{N^\sigma}$. The leading term $\omega^N$ is characterized by the value of the connectivity constant $\omega>1$, which depends on $b$, but not on the type of HW. In contrast to that, the stretched exponential term $\mu^{N^\sigma}$ depends on the type of HW through constant $\mu<1$, whereas exponent $\sigma$ is determined by $b$ alone. For larger $b$ values, using some general features of the applied recursive relations, without explicit enumeration of HWs, we argue that asymptotical behavior of $Z_N$ should be the same, with $\sigma=\ln 3/\ln[b(b+1)(b+2)/6]$, valid for all $b>2$. This differs from the formulae obtained recently for Hamiltonian walks  on other fractal lattices, as well as from the formula expected for homogeneous lattices. We discuss the possible origin and implications of such result. \end{abstract}


\section{Introduction}

The fact that compact conformations of polymers are principal configurations
of the native states of globular proteins makes them an important subset of all the physically
accessible conformations. Since compact conformations occupy space as densely as possible, one of the simplest ways to model them is to use Hamiltonian walks (HWs) on a lattice, which are, by definition, self-avoiding walks (SAWs) that visit all the lattice sites exactly once \cite{Vanderzande}. In order to make this model more capable for capturing
different features of various physical phenomena (such as protein melting \cite{Flory} or protein folding \cite{Dill}) local interactions can be introduced, but even in its simplest form, with no interactions taken into account, the problem of enumeration and classification of Hamiltonian walks proved to be extremely difficult.

Studies of Hamiltonian walks are primarily focused on finding the overall numbers $Z_N$ of open and closed HWs on lattice with $N$ sites. It is expected that limiting value of ${\ln Z_N}/{N}$  exists when $N\to\infty$, and its particular value
\[
\ln\omega=\lim\limits_{N\to\infty}\frac{\ln Z_N}{N}
\]
corresponds to the configurational entropy per monomer (site). This means that to the lowest approximation,
$Z_N$ behaves as $\omega^N$ and, therefore, the so-called connectivity constant $\omega>1$ can be interpreted as average number of steps available to the walker having already completed a large number of steps. The leading corrections to the exponential term are expected to have the power-law form $N^a$ (as in the case of ordinary self-avoiding walks), and the stretched exponential form $\mu^{N^\sigma}$, with $\mu<1$,  so that $Z_N$ should scale as
\begin{equation}
Z_N\sim \omega^N {\mu^{N^\sigma}}N^a\, .
\label{eq:asimptotika}
\end{equation}
Expectations of such forms are based on the exact studies of HWs on the Manhattan \cite{Manhattan} and some fractal lattices \cite{Bradley,Stajic,Elezovic}, as well as on results obtained for closely related models of collapsed interacting self-avoiding walks on square \cite{Prellberg,SamoOwczarek,BennetWood,Baiesi} and cubic lattice \cite{Grassberger}.
In the case of collapsed SAWs appearance of the stretched exponential term, which is not present in the scaling form
for non-interacting SAWs, was explained as a consequence of surface effects. Namely, a collapsed SAW forms a compact globule, with a sharp boundary separating it from the surrounding solvent, so that monomers on the boundary have smaller number of contacts with other monomers than those in the bulk of the globule, and therefore surface tension should arise. For homogeneous lattices one can assume that the boundary itself is homogeneous surface, and then straightforward arguments \cite{Owczarek} lead to the conclusion that term  $\mu^{N^\sigma}$,  with $\sigma=(d-1)/d$ ($d$ being the dimensionality of the lattice), should appear. Exact studies \cite{Samo Owczarek},
series analysis of data obtained via exact enumeration \cite{Prellberg,BennetWood}, as well as Monte-Carlo simulations \cite{Baiesi,Grassberger} of collapsed SAWs on homogeneous lattices, certainly confirm existence of the stretched exponential term with the proposed formula for $\sigma$. Although it is believed that HW model corresponds to the interacting SAW model at temperature $T=0$, in spite of the continuous improvements of exact  \cite{Mayer} and Monte Carlo  \cite{Ramakrishnan} enumeration techniques, direct confirmation of scaling relation \eref{eq:asimptotika} for HWs on non-oriented homogeneous lattices has not been achieved yet. Apart from Manhattan lattice, scaling forms for Hamiltonian walks were obtained only for some fractal lattices, where stretched exponential corrections for both open and closed HWs  were found only for $n$-simplex fractals with even $n$ \cite{Elezovic}. Even in that cases, its presence cannot be explained using a simple generalization of the argument used for collapsed self-avoiding walks on homogeneous lattices. However, results of the studies on fractals suggest that stretched exponential term can be expected on lattices where larger number of entangled conformations is possible. In order to get a deeper insight into this issue, here we apply an exact recursive method for enumeration and classification of HWs on modified three-dimensional Sierpinski gasket family of fractals. In contrast to most of the previously in this context studied fractals, which were embedded either in two-dimensional spaces or spaces with dimensionality higher than 3, each member of the family studied in the present paper is embedded in three-dimensional space. Therefore, HWs on these fractals can be understood as a toy model for compact polymers critical behaviour in realistic nonhomogeneous 3d media, which, to the best of our knowledge, has not been studied so far.

The paper is organized as follows. Modified 3d Sierpinski gasket fractals, as well as the method itself, are described in the next section. Explicit forms of the recursion relations, obtained for the types of Hamiltonian walks which are needed for generation and enumeration of all closed conformations are presented for the first three members of the fractal family. Numerically analyzing these relations we find scaling form $\omega^N \mu_C^{N^\sigma}$. Using some general features of these relations (obtained in \ref{ap:kb} and \ref{app:xasimptotika}), we argue that such form should be correct for the whole fractal family, and derive the closed-form formula for the exponent $\sigma$. In section 3 we extend the method to open HWs, explicitly apply it again on the first three fractals, and then generalize it. It turns out that the number of open HWs scales as $\omega^N \mu_O^{N^\sigma}$, with the same values of the connectivity constant $\omega$ and the exponent $\sigma$ as for the closed HWs, but with $\mu_O\neq \mu_C$. General features of the recursion relations for open HWs are derived in \ref{app:ac} and \ref{app:dh}. All the obtained results are summarized and discussed in the section 4, and some auxiliary considerations, needed for the discussion, are given in Appendix E.

\section{Closed Hamiltonian walks on modified three-dimensional Sierpinski gaskets  \label{sec:zatvorene}}

\begin{figure}
\begin{center}
\includegraphics[height=.35\textheight]{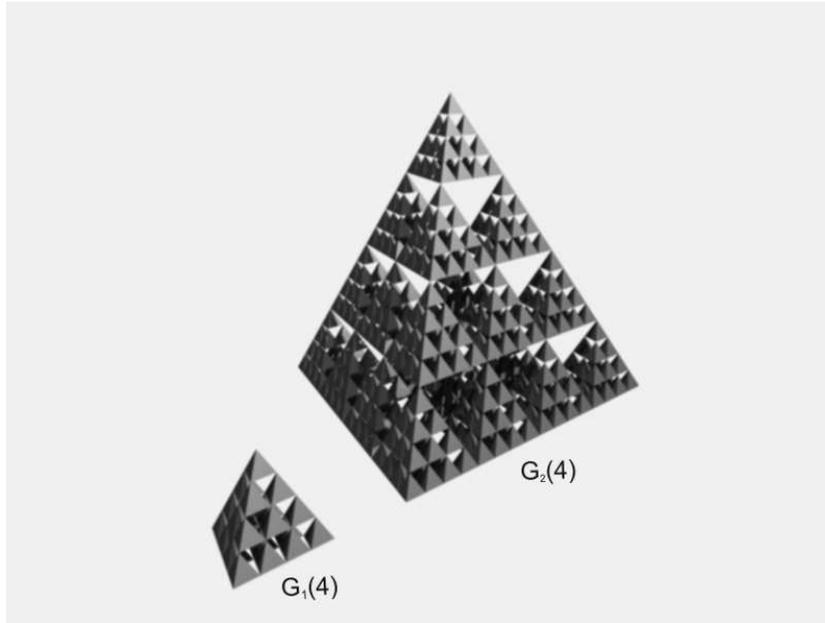} \end{center}
\caption{The first two steps of the construction of the 3d MSG fractal with $b = 4$. Vertices of neighboring unit tetrahedrons are connected via infinitesimal junctions.}\label{fig:fraktal}
\end{figure}

Three-dimensional modified Sierpinski gasket (3d MSG) fractal is constructed recursively, starting with a unit tetrahedron. The first step of the
construction, so called generator $G_1(b)$ of order $l=1$,  is obtained by joining
\begin{equation}
N_G=b(b+1)(b+2)/6  \label{eq:NG}
\end{equation}
unit tetrahedrons into a
$b$ times larger tetrahedral structure (see \fref{fig:fraktal}) in such a way that vertices of neighboring unit
tetrahedrons are connected via infinitesimal junctions.
Enlarging the generator $b$ times, and substituting the smallest
tetrahedrons with $G_1(b)$, and then repeating this procedure $l$ times, one obtains $G_{l+1}(b)$ - generator of order
$l+1$, which contains $N_{l+1}=4N_G^{l+1}$ sites. The complete 3d MSG fractal with parameter $b$ is obtained when $l\to\infty$, and its fractal dimension $d_f$ is equal to
\begin{equation}
 d_f=\ln [b(b+1)(b+2)/6]/{\ln b}\, .
 \end{equation}
One should note here that ordinary 3d Sierpinski gasket (SG) fractal with parameter $b$ is constructed in a similar way, with the only difference being that neighboring unit tetrahedrons in 3d SG lattice are not moved apart, as is the case for the 3d MSG fractal. This small change does not alter the basic properties of the lattice, such as its fractal dimension, but it simplifies the complicated scheme needed for enumeration of closed Hamiltonian walks on 3d SG fractals \cite{Stajic} and allows for enumeration of extremely long open HWs on 3d MSG fractals with $b=2,3$, and 4. In addition, general conclusions about scaling forms of the numbers of HWs on 3d MSG fractal with arbitrary $b>2$ can be derived, as will be explained in what follows.

\begin{figure} \begin{center} \includegraphics[height=.35\textheight]{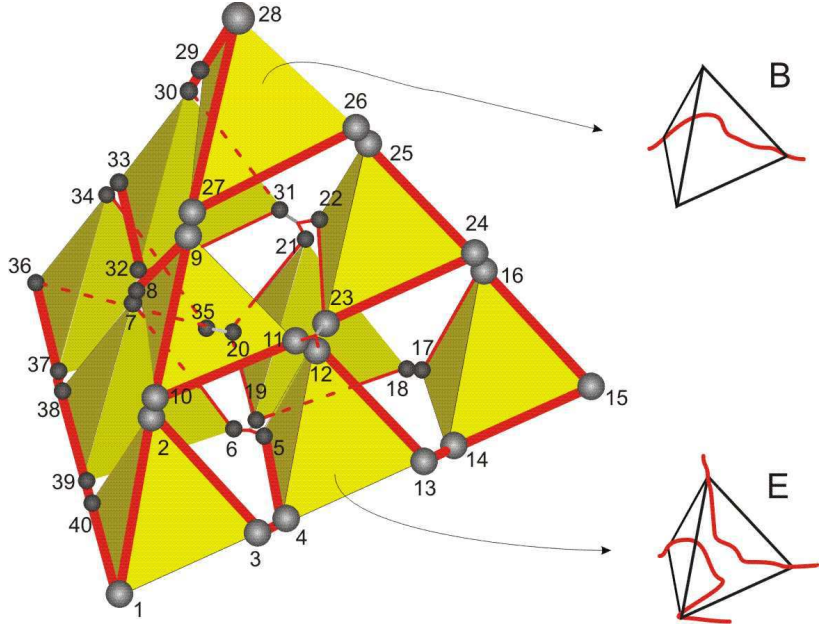}
\end{center} \caption{Example of closed Hamiltonian walk (red line) on the generator of the 3d $b=3$ MSG fractal. The generator
consists of 40 vertices, which are numbered consecutively from the first (arbitrarily chosen) vertex visited by the HW, to the last one (which is connected with the first one). One can notice that HW conformation within the unit
tetrahedron can be either one-stranded ($B$-type), like the one connecting the sites 26-27-28-29, or two-stranded ($E$-type), like the conformation
traversing the tetrahedron with vertices 4-5-12-13.} \label{fig:primerZatvorena}
\end{figure}

By definition, each Hamiltonian walk on $G_l(b)$ structure visits all its $N_l$ sites exactly once.
In \fref{fig:primerZatvorena}
an example of closed Hamiltonian walk on the
generator $G_1(3)$ is shown. One can notice that this HW can be decomposed into $N_G=10$ parts (within the ten unit
tetrahedrons), which are either one- or two-stranded. This observation can easily be generalized to generators of
higher
order and any $b$: any closed HW on $G_{l+1}(b)$ can be decomposed into $N_G$ parts within the same number of
generators $G_l(b)$, from which the generator $G_{l+1}(b)$ is made of. These parts traverse each
$G_{l}(b)$ either once or two times, and we shall call the corresponding configurations $B$- or $E$-type HW steps,
respectively, whereas numbers of HWs of these types within a $G_{l}(b)$ will be denoted by $B_l$ and $E_l$,
respectively. Using these numbers, the overall number of closed HWs on $G_{l+1}(b)$ can be expressed as
\begin{equation}
Z_{l+1}^C=\sum\limits_{k=0}^{k_C}n_kB_l^{N_G-k}E_l^k\, , \label{eq:zatvorene}
\end{equation}
where $n_k$ are the numbers of closed HWs configurations within $G_{l+1}(b)$, with $k$ steps of $E$--type, and $(N_G-k)$ $B$--steps. For instance, the path presented in \fref{fig:primerZatvorena} is one of $Z_1^C$ possible closed HWs within the $G_1(3)$, contributing to the term  $n_2B_0^8E_0^2$ in the corresponding equation \eref{eq:zatvorene}. The upper limit $k_C$ in the sum in \eref{eq:zatvorene} is equal to 0 for $b=2$, whereas for $b>2$ it can be shown (see \ref{ap:kb}) that
\begin{equation}
k_C=\frac 12(b+1)(b+2)-8\, . \label{eq:kc}
\end{equation}
Due to the self-similarity of the lattices under study, the numbers $n_k$ do not depend on $l$, and, also,  numbers $B_l$ and $E_l$ fulfil recursion relations of the following form:
\begin{equation}
B_{l+1}=\sum\limits_{k=k_B}^{N_G}m_kB_l^{k}E_l^{N_G-k}\, , \quad
E_{l+1}=\sum\limits_{k=k_E}^{N_G}p_kB_l^kE_l^{N_G-k}\, , \label{eq:eb}
\end{equation}
where coefficients $m_k$ and $p_k$ depend only on $k$ and $b$, and are positive integers for all $b>2$, including zero for $b=2$.
For instance, for $b=2$ these relations are\footnote{Note that for
$b=2$ recursion relations \eref{eq:eb}  are the same as for the 4-simplex fractal lattice \cite{Bradley}.}
\begin{equation} B_{l+1}=2B_l^4+4B_l^3E_l+6B_l^2E_l^2\, , \quad E_{l+1}=B_l^4+4B_l^3E_l+22E_l^4\, , \label{eq:rrb2}
\end{equation}
and we were able to find the explicit form of relations \eref{eq:eb} for $b=3$ and 4 also, by direct computer enumeration of possible HW conformations within the MSG generator. The corresponding coefficients $m_k$ and $p_k$ are presented in
\tref{tab:rec}, \begin{table}
\caption{ \label{tab:rec} Coefficients appearing in recursion relations (\ref{eq:eb}), found
by direct computer enumeration of the corresponding HW conformations on 3d MSG fractals with $b=2,3,4$.}
\begin{indented}
\item[]
\begin{tabular}{@{}ccccccccccccc} \br \multicolumn{3}{c}{$b=2$} &&&\multicolumn{3}{c}{$b=3$}
&&&\multicolumn{3}{c}{$b=4$} \\ \mr $k$&$m_k$&$p_k$&&&$k$&$m_k$&$p_k$&&&$k$&$m_k$&$p_k$\\ \mr
0&-&22&&&2&-&4308&&&7&-&26465392\\ 1&-&0&&&3&-&1936&&&8&-&99652120\\ 2&6&0&&&4&-&5808&&&9&-&151443088\\
3&4&4&&&5&3192&1888&&&10&23848720&199987864\\ 4&2&1&&&6&848&2534&&&11&58605536&204194352\\
-&-&-&&&7&1728&1056&&&12&78351952&172479256\\ -&-&-&&&8&664&596&&&13&81469824&126633376\\
-&-&-&&&9&332&160&&&14&66418856&78454776\\ -&-&-&&&10&64&32&&&15&43526336&41200784\\
-&-&-&&&-&-&-&&&16&22989024&18548660\\ -&-&-&&&-&-&-&&&17&9642816&6662824\\ -&-&-&&&-&-&-&&&18&3032724&1901008\\
-&-&-&&&-&-&-&&&19&626056&397392\\ -&-&-&&&-&-&-&&&20&62434&42514\\
\br
\end{tabular}
\end{indented}
\end{table}
while for larger $b$ values they could not be reached within the reasonable time  with the computer facilities available to us. However, one can show (see \ref{ap:kb}) that for general $b>2$ lower limits of the sums in relations \eref{eq:eb} are equal to
\begin{equation}
 k_B=N_G-k_C-3=\frac 16(b+1)(b+2)(b-3)+5\, , \qquad k_E=k_B-3  \, ,
\label{eq:kbke}
\end{equation}
which is a result essential for establishing some general conclusions, as will be explained in the
following paragraphes.

Once the explicit form of recursion relations \eref{eq:zatvorene} and \eref{eq:eb} is established, starting with the
initial values $B_0=2$, and $E_0=1$, corresponding to the unit tetrahedron, one can calculate the numbers $B_l$, $E_l$, and subsequently the overall number $Z_{l+1}^C$ of closed HWs, in principle for any $l$. However, since these numbers quickly become extremely large, it is useful to introduce new variable $x_l=B_l/E_l$. Then, from \eref{eq:eb} it follows that  recursion relation for the variable $x_l$, for $b>2$\footnote{The corresponding analysis of the $b=2$ case is the same as for 4-simplex fractal lattice, which is given in \cite{Bradley}.}, has the form
\begin{equation}
x_{l+1}=x_l^{k_B-k_E}g(x_l)=x_l^{3}g(x_l)\, ,
\qquad g(x)=\frac{\sum\limits_{k=0}^{k_C+3}m_{k_B+k}\, x^k}{\sum\limits_{k=0}^{k_C+6}p_{k_E+k}\, x^k}\, .
\label{eq:xl}
\end{equation}
Numerical analysis of this recursion relation for $b=3$ and $b=4$ cases, shows that
starting with the initial value $x_0=2$,  $x_l$ quickly tends to 0, as $l$ grows. Assuming that this is correct for
general $b>2$, for large values of $l$ the above equation can be approximated as
\begin{equation}
x_{l+1}\approx
\frac{m_{k_B}}{p_{k_E}}\, x_l^3\, , \label{eq:xlpriblizno}
\end{equation}
from which follows
\begin{equation} x_l\sim \lambda^{3^l}\, ,
\label{eq:asimptotikaxl}
\end{equation}
where $\lambda$ is some constant, whose value depends on $b$, but it is always
less than 1.  For $b=2$ the corresponding relation is $x_l\sim \lambda^{2^l}$ \cite{Bradley}, and particular values of
$\lambda$ for $b=2,3$ and 4 cases are given in \tref{tab:rezhw}.\begin{table} \caption{\label{tab:rezhw} Values of
relevant constants appearing in the scaling forms (\ref{eq:scalingZatvorene}) and (\ref{eq:scalingOtvorene}) of the
overall numbers of HWs on 3d MSG fractals, with $b=2,3,4$, together with the corresponding values of the fractal
dimension $d_f$. } \begin{indented} \item[] \begin{tabular}{@{}crcllllll} \br
{$b$}&{$d_f$}&{$k_C$}&{$\lambda$}&{$\omega$} &{$\mu_C$} &{$\mu_O$}&{$\sigma$}&{$1/d_f$} \\ \mr 2& ${\ln 4}/{\ln
2}$&0&0.8366&1.0876&0.8366&0.9147&0.5&0.5 \\ 3& ${\ln 10}/{\ln 3}$&2&0.8835&1.4404&0.8963&0.9554&0.4471&0.4471\\ 4&
${\ln 20}/{\ln 4}$&7&0.8639&1.4686&0.8696&0.9496&0.3667&0.4628\\ \br \end{tabular} \end{indented} \end{table}

Recursion relation for numbers $E_l$ of two-stranded HWs \eref{eq:eb}, with the variable $x_l=B_l/E_l$
obtains the form
\begin{equation}
 E_{l+1}=E_l^{N_G}x_l^{k_E}f(x_l)\, ,\qquad f(x)=\sum\limits_{k=0}^{k_C+6}p_{k+k_E}x^k\, , \label{eq:funkcijaf}
\end{equation}
from which one gets
\begin{equation}
\frac{\ln E_{l+1}}{4N_G^{l+1}}=\frac{\ln
E_{l}}{4N_G^{l}}+k_E\frac{\ln x_l}{4N_G^{l+1}}+\frac{\ln f(x_l)}{4N_G^{l+1}}\, .  \label{eq:elrek}
\end{equation}
Numerically iterating this recursion relation, together with \eref{eq:xl}, one finds that
\begin{equation}
\lim_{l\to\infty}\frac{\ln E_{l}}{4N_G^{l}}=\ln\omega\, ,  \label{eq:limel}
\end{equation}
where $\omega$ (see \tref{tab:rezhw}) is constant larger than 1. On the other hand, the overall number $Z_{l+1}^C$ \eref{eq:zatvorene} of closed HWs on the generator of order $l+1$ can be expressed as
\begin{equation}
Z_{l+1}^C=E_l^{N_G}x_l^{N_G-k_C}h(x_l)\, ,\qquad h(x)=\sum\limits_{k=0}^{k_C}n_{k_C-k}\,x^k\,,
\label{eq:zatvPomoc}
\end{equation}
so that
\begin{equation} \frac{\ln Z^C_{l+1}}{4N_G^{l+1}}=\frac{\ln
E_{l}}{4N_G^{l}}+\frac{N_G-k_C}{4N_G}\frac{\ln x_l}{N_G^l}+\frac{\ln h(x_l)}{4N_G^{l+1}}\, . \label{eq:zcrel}
\end{equation}
From the asymptotical behavior \eref{eq:asimptotikaxl} of the number $x_l$, and from the fact that
$h(x)$ tends to constant value $n_{k_C}>0$ when $x\to 0$, it then follows that
\[
\lim_{l\to\infty}\frac{\ln
Z_{l}^C}{N_{l}}=\lim_{l\to\infty}\frac{\ln E_{l}}{4N_G^{l}}=\ln\omega\, ,
\]
where $N_l=4N_G^l$ is overall number of vertices within the generator of order $l$.

To find the leading-order correction to the asymptotic behavior of $Z_l^{C}$ we first introduce variable
\begin{equation}
y_l=\frac{\ln E_{l}}{4N_G^{l}}=\frac{\ln E_{l}}{N_l}\, , \label{eq:yl}
\end{equation}
which, as follows from \eref{eq:elrek}, satisfies the relation
\[
y_l=\sum_{k=0}^{l-1}(y_{k+1}-y_k)+y_0=\sum_{k=0}^{l-1}\frac
1{N_{k+1}}\left[k_E\ln x_k+\ln f(x_k)\right]\, .
\]
Then, using \eref{eq:limel}, one obtains
\[
y_l=\ln\omega-\sum_{k=l}^{\infty}\frac 1{N_{k+1}}\left[k_E\ln x_k+\ln f(x_k)\right]\, ,
\]
from which, taking into account the large $k$ behavior \eref{eq:asimptotikaxl} of $x_k$, it follows that
\[
y_l\approx \ln\omega
-\frac 1{N_l}\frac{k_E}{N_G-3}\, 3^l\, \ln\lambda-\frac{\mathrm{const}}{N_l}\, , \qquad l\gg 1\, .
\]
Substituting this relation into \eref{eq:zcrel} one derives
\begin{equation}
\ln Z_l^C\approx N_l\ln\omega+N_l^{\sigma}\ln\mu_C\, ,\quad i.e. \quad Z_l^C\sim \omega^{N_l}\mu_C^{N_l^\sigma}\, ,
\label{eq:scalingZatvorene}
\end{equation}
with
\begin{equation}
\sigma=\frac{\ln 3}{\ln N_G}=\frac{\ln 3}{\ln
[b(b+1)(b+2)/6]}\, ,   \label{eq:sigma}
\end{equation}
and
\begin{equation}
\mu_C= \lambda^A\, , \qquad     A=\frac{N_G+k_C}{N_G-3}\,4^{-\sigma}\, .\label{eq:muc}
\end{equation}

At the end of this section we want to emphasize that the crucial step in preceding derivation, which led to the scaling form  \eref{eq:scalingZatvorene}, with the value of exponent $\sigma$ given by formula \eref{eq:sigma}, was the assumption that $x_l$ tends to 0 for general $b$ (explicitly confirmed only up to $b=4$).
Its direct consequence is approximate difference equation \eref{eq:xlpriblizno}, which is obtained from the exact relation \eref{eq:xl}. On the other hand, the key ingredient of that relation is the fact that $k_B-k_E=3$, due to which $x_l$ behaves as $\lambda^{3^l}$, for $l\gg 1$. In that sense, exact expressions for $k_B$ and $k_E$ \eref{eq:kbke}, obtained in Appendix A for general $b>2$, are essential for establishing the scaling form \eref{eq:scalingZatvorene}, whereas the particular values of the coefficients $m_k$ and $p_k$ of the recursion relations \eref{eq:eb} do not affect neither its general form, nor the value of $\sigma$. However, whether $x_l$ tends to 0 or not certainly depends on the values of $m_k$ and $p_k$. Analyzing data given in \tref{tab:rec} for $b=3$ and $b=4$, one can observe that $p_{k_E+k}>m_{k_E+k+1}$, for $2\leq k\leq k_C+5$. It can be shown (see \ref{app:xasimptotika}) that such inequality for general $b>2$ is sufficient to prove that numbers $x_l$, which satisfy difference equation \eref{eq:xl} with the initial condition $x_0=2$, tend to 0 when $l\to\infty$. Unfortunately, the inequality itself we were not able to prove, but, since it seems plausible, we think that scaling relation \eref{eq:scalingZatvorene} can be accepted as valid for general $b$.

\section{Open Hamiltonian walks on modified three-dimensional Sierpinski gaskets}

\begin{figure}
\begin{center}
\includegraphics[height=.5\textheight]{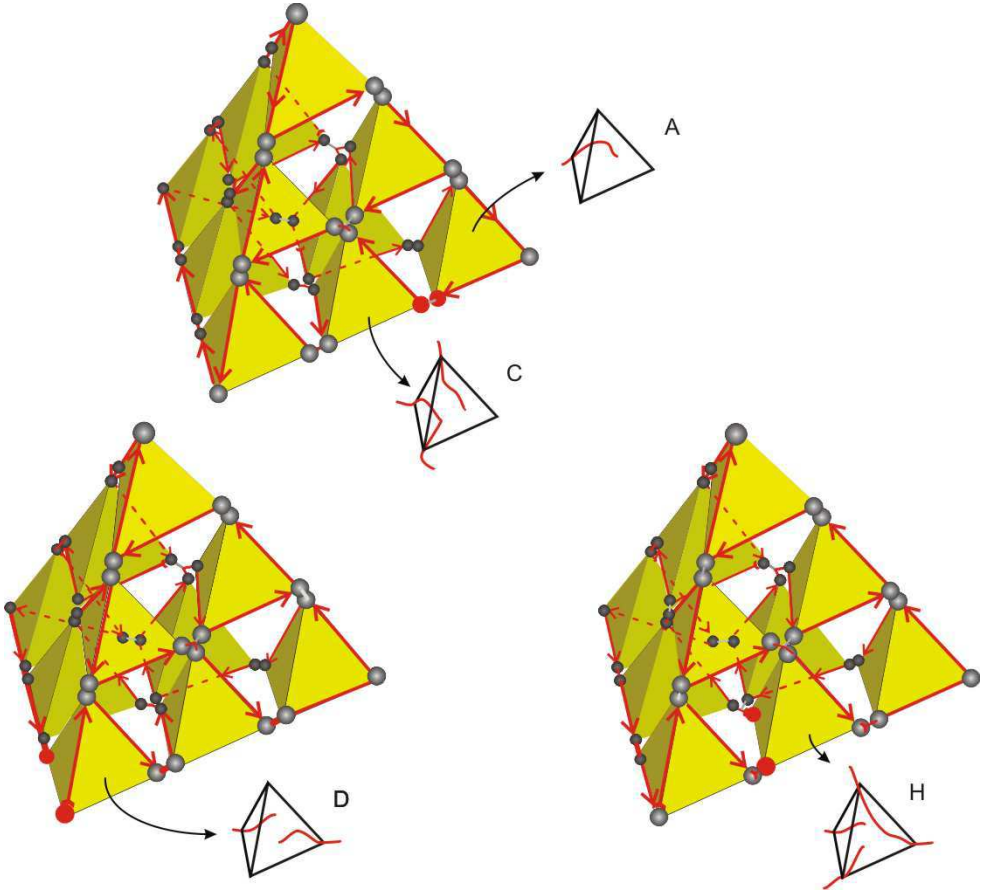}
\end{center}
\caption{Three examples of open HW (oriented red lines, where the only meaning of arrows is to serve as guides to the
eye) on the generator of the 3d $b=3$ MSG fractal. Points at which HW begins or terminates are marked as full red
circles, for the sake of better recognition. If both ends of the walk are in the same tetrahedron the two-leg HW
conformation within that tetrahedron is either of $D$- or $H$-type. Otherwise, if end-points of the walk belong to
different tetrahedrons, the corresponding one-leg conformations are of $A$- or $C$-type.} \label{fig:primerOtvorena}
\end{figure}

Any open HW on generator $G_{l+1}(b)$ of order $l+1$ can be decomposed into $N_G$ parts within its $N_G$ constitutive
generators $G_l(b)$ of order $l$. The parts which contain ending points (see \fref{fig:primerOtvorena}) can be of four
different types:
\begin{itemize}
\item $A$-type, which consists of one HW, with one end at one vertex of $G_{l}(b)$ and
the other end at any other  site of $G_l(b)$, including its vertices;
\item $C$-type, composed of two non-intersecting SAWs, one with both ends at the vertices of the $G_l(b)$, and the
    other with one end at the third vertex of the $G_l(b)$ and the second anywhere within it - these two SAWs
    together visit all the sites of the $G_l(b)$;
\item $D$-type, consisting of two non-intersecting SAWs, each of them starting at different vertex of $G_l(b)$ and
    ending anywhere within it, in such a way that all sites of $G_l(b)$ are visited;
\item $H$-type, comprised of three non-intersecting SAWs, each of them starting at different vertex of $G_l(b)$, one
    ending at the fourth vertex, and the remaining two anywhere within $G_l(b)$ - again, all sites of $G_l(b)$
    should be visited by these three SAWs.
\end{itemize}
$A$-- and $C$--type configurations have one dangling end, therefore we shall call them one-leg
configurations (steps), whereas $D$-- and $H$--configurations, with two dangling ends, will be called two-leg steps. If both end-points of the complete HW lie in the same $G_l(b)$, that $G_l(b)$ contains a two-leg step (of $D$-- or $H$--type), whereas the remaining $N_G-1$ generators $G_l(b)$ contain either $B$-- or $E$--step. Otherwise, when end-points are in two different $G_l(b)$ generators, the complete open HW has two one-leg steps (of $A$-- or $C$--type), and the remaining $N_G-2$ parts of the walk are of either $B$-- or $E$--type. Therefore, one concludes that overall number $Z_{l+1}^O$ of open HWs  within $G_{l+1}(b)$ is equal to
\begin{equation}
Z_{l+1}^O=A_l^2 F_{AA}+A_lC_lF_{AC}+C_l^2F_{CC}+D_lS_D +H_lS_H\, ,  \label{eq:otvorene}
\end{equation}
where $F_{AA}$, $F_{AC}$, $F_{CC}$, $S_D$ and $S_H$ are polynomials in $B_l$ and $E_l$, of power $N_G-2$ or $N_G-1$,  and $A_l$, $C_l$, $D_l$ and $H_l$ are overall numbers of corresponding type HWs within $G_l(b)$ generator. Numbers of one-leg conformations fulfill recursion relations of the following form:
\begin{equation}
\eqalign{
A_{l+1}=R_{11}(B_l,E_l)A_{l}+R_{12}(B_l,E_l)C_{l}\, ,\\
C_{l+1}=R_{21}(B_l,E_l)A_{l}+R_{22}(B_l,E_l)C_{l}\, ,}
\label{eq:ac}
\end{equation}
with
\begin{equation}
R_{ij}(B,E)=\sum\limits_{k=0}^{k_{ij}}r^{ij}_kB^{N_G-1-k}E^k \, ,
\label{eq:r}
\end{equation}
while numbers of two-leg conformations obey relations of the form
\begin{equation}
\eqalign{
D_{l+1}=d_{D}D_l+d_{H}H_l+d_{AA}A_l^2+d_{AC}A_lC_l+d_{CC}C_l^2\, ,   \\
H_{l+1}=h_{D}D_l+h_{H}H_l+h_{AA}A_l^2+h_{AC}A_lC_l+h_{CC}C_l^2\, ,}  \label{eq:h}
\end{equation}
with $d$ and $h$ being some polynomials in $B_l$ and $E_l$. For instance, for $b=2$ relations \eref{eq:otvorene} and
\eref{eq:ac} are:
\begin{eqnarray}
Z_{l+1}^O=12B_l^2(A_l^2+2A_lC_l+3C_l^2+B_lD_l)\, , \nonumber\\
A_{l+1}=(6B_l^3+6B_l^2E_l)A_l+(12B_l^3+18B_l^2E_l)C_l\, , \nonumber\\
C_{l+1}=(B_l^3+3B_l^2E_l)A_l+(3B_l^3+12B_l^2E_l+16B_lE_l^2+16E_l^3)C_l\, .\nonumber
\end{eqnarray}
They coincide with the corresponding relations for 4-simplex fractal lattice, as well as does the complete further
analysis, which can be seen in \cite{Elezovic}.  In the remaining part of this section we present the general analysis
for $b>2$ 3d MSG fractals.

For general $b>2$ it can be shown (see \ref{app:ac}) that polynomials $F$ and $S$, appearing in \eref{eq:otvorene}, have the form
\begin{equation}
\eqalign{
F_{XY}(B,E)=\sum_{k=0}^{k_C+1}z^{XY}_k B^{N_G-k-2}E^k\, ,  \quad XY=AA, AC, CC  \\
S_D(B,E)=\sum_{k=0}^{k_C}z^D_k B^{N_G-k-1}E^k\, , \quad  S_H(B,E)=\sum_{k=0}^{k_C-1}z^H_k B^{N_G-k-1}E^k\, .}
\label{eq:fxy}
\end{equation}
Starting with the initial values for the numbers $B_l$, $E_l$, $A_l$, $C_l$, $D_l$, and $H_l$, and using the recursive relations \eref{eq:eb}, \eref{eq:ac}, and \eref{eq:h},  one can calculate these numbers in principle for any $l$, and substituting them into \eref{eq:otvorene}, eventually find the overall number $Z_{l+1}^O$ of open HWs. Since all these numbers increase rapidly with $l$,  it is useful to introduce variables
\begin{equation}
u_l=\frac{A_l}{E_l}\, , \qquad v_l=\frac{C_l}{E_l}\, , \qquad w_l=\frac{D_l}{E_l}\, , \qquad q_l=\frac{H_l}{E_l}\, ,
\end{equation}
in addition to already defined $x_l=B_l/E_l$. With these variables $Z_{l+1}^O$ can be rewritten as
\begin{eqnarray}
\fl
Z_{l+1}^O=E_l^{N_G}x_l^{N_G-k_C-3}\left[\sum\limits_{k=0}^{k_C+1}\left(z_{k_C+1-k}^{AA}u_l^2+z_{k_C+1-k}^{AC}u_lv_l+
z_{k_C+1-k}^{CC}v_l^2\right)x_l^k\right.\nonumber\\ +\left. w_lx_l^2\sum\limits_{k=0}^{k_C}z_{k_C-k}^Dx_l^k+
q_lx_l^3\sum\limits_{k=0}^{k_C-1}z_{k_C-k-1}^Hx_l^k\right]\, , \label{eq:otvoreneuv}
\end{eqnarray}
and explicitly calculated using corresponding recursion relations for $u_l$, $v_l$, $w_l$, and $q_l$, simultaneously with \eref{eq:xl} and \eref{eq:funkcijaf}. In \ref{app:ac} it is shown that the upper limits of the sums in \eref{eq:r} are equal to
\begin{equation}
k_{11}=k_{12}=k_C+2\, , \qquad  k_{21}=k_{22}=k_C+5\, ,  \label{eq:rgranice}
\end{equation}
so that by dividing \eref{eq:ac} with recursion relation for $E_l$, given in \eref{eq:eb}, for $u_l$ and $v_l$ one
obtains recursion relations
\begin{equation}
\left(
  \begin{array}{c}
    u_{l+1} \\
    v_{l+1} \\
  \end{array}
\right)=\left(
          \begin{array}{cc}
            m_{11} & m_{12} \\
            m_{21} & m_{22} \\
          \end{array}
        \right)\left(
  \begin{array}{c}
    u_{l} \\
    v_{l} \\
  \end{array}
\right)\, ,  \label{eq:yz}
\end{equation}
where $m_{ij}$ are functions of $x_l$ of the following form:
\[
\fl m_{1i}(x)=\frac{x^{3}}{f(x)}\sum_{k=0}^{k_C+2}r_{k_C+2-k}^{1i}  x^k\, , \quad
m_{2i}(x)=\frac
1{f(x)}\sum_{k=0}^{k_C+5}r_{k_C+5-k}^{2i} x^k\, , \quad i=1,2\, ,
\]
and $f(x)$ is defined in \eref{eq:funkcijaf}. Since $x_l\to 0$ and $f(x_l)\to\mathrm{const}\neq 0$ when $l\to\infty$, it follows that $m_{1i}(x_l)\to 0$, $m_{2i}(x_l)\to\mathrm{const}\neq 0$, implying that $u_l$ tends to 0, and $v_l$ to some constant value. This is indeed correct for $b=3$ and $b=4$ cases, for which we managed to find the complete set of coefficient $r_k^{ij}$ (see \tref{tab:opencoef}).
\begin{table}
\caption{\label{tab:opencoef}Coefficients of the
polynomials $R_{ij}$ \eref{eq:r}, appearing in recursion relations \eref{eq:ac}, found by direct computer enumeration
of the corresponding HW conformations on 3d MSG fractals with $b=3$ and 4.}
\raggedleft
\footnotesize\rm
\begin{tabular}{@{}cccccccccc} \br  &\centre{4}{$b=3$} &\centre{4}{$b=4$} \\  &\crule{4}&\crule{4}\\
$k$&$r^{11}_k$&$r^{12}_k$&$r^{21}_k$&$r^{22}_k$ &$r^{11}_k$&$r^{12}_k$&$r^{21}_k$&$r^{22}_k$\\
 \mr
0&378&1050&202&620&726000&2078460&454340&1343670\\ 1&1722&5256&798&2854&6785652&20162052&4124238&12825924\\
2&2748&8964&1962&8696&30159948&92803776&18522682&61201186\\
3&3588&15144&2912&15172&85107696&270289488&56914854&200289608\\
4&2976&9576&3126&24064&171919920&559983864&135608360&503818972\\
5&-&-&4632&23684&265804824&875876784&260879696&1011455276\\
6&-&-&2420&27088&319608912&1041132024&412200376&1653387800\\
7&-&-&1304&4308&290292720&895898352&548029844&2243585880\\ 8&-&-&-&-&185693208&493595664&612939852&2479455304\\
9&-&-&-&-&51662736&71546160&552106184&1451395108\\ 10&-&-&-&-&-&-&407928692&1451395108\\
11&-&-&-&-&-&-&226652724&546796688\\ 12&-&-&-&-&-&-&68954256&26465392\\
 \br
\end{tabular}
\end{table}
Starting with the initial values $A_0=6$, $C_0=2$, {\em i.e.} $u_0=6$ and $v_0=2$, and numerically iterating relations \eref{eq:yz}, for both $b=3$ and $b=4$, already after five iterations one obtains limiting values of $v_l$: $441.32\ldots$ for $b=3$, and $3538.91\ldots$ for $b=4$.

In \ref{app:dh} it is shown that recursion relations for the variables $w_l$ and $q_l$ can be put into the following matrix form
 \begin{equation}
 \left(
  \begin{array}{c}
    w_{l+1} \\
    q_{l+1} \\
  \end{array}
\right)=\left(
          \begin{array}{cc}
            p_{11} & p_{12} \\
            p_{21} & p_{22} \\
          \end{array}
        \right)\left(
  \begin{array}{c}
    w_{l} \\
    q_{l} \\
  \end{array}
\right)+\left(
  \begin{array}{c}
    t_{11}u_l^2+t_{12}u_lv_l+t_{13}v_l^2 \\
    t_{21}u_l^2+t_{22}u_lv_l+t_{23}v_l^2\\
  \end{array}
\right)\, , \label{eq:wq}
\end{equation}
where $p_{ij}$ and $t_{ij}$ are functions of $x_l$ of the form
\begin{eqnarray}
\fl p_{11}(x)&=&\frac{x^2}{f(x)}{\sum\limits_{k=0}^{k_C+3}(k+k_B)m_{k+k_B}x^k}\, , \quad
 p_{12}(x)=\frac{2x^3}{f(x)}\sum\limits_{k=0}^{k_C+2}(k_C+3-k)m_{k+k_B}x^k\, ,\nonumber\\
\fl p_{21}(x)&=&\frac{1}{2xf(x)}\sum\limits_{k=0}^{k_C+6}(k+k_E)p_{k+k_E}x^k\, ,\quad
p_{22}(x)=\frac{1}{f(x)}\sum\limits_{k=0}^{k_C+6}(k_C+6-k)p_{k+k_E}x^k\, ,\nonumber\\
\fl t_{1i}(x)&=&\frac 1{f(x)}{\sum_{k=0}^{k_C+4}s^{XY}_{k_C+4-k}x^k}\, ,\quad
t_{21}(x)=\frac 1{x^{3}f(x)}{\sum\limits_{k=0}^{k_C+7}o^{AA}_{k_{C}+7-k}x^k}\, ,\nonumber\\
\fl t_{22}(x)&=&\frac 1{x^{2}f(x)}{\sum\limits_{k=0}^{k_C+6}o^{AC}_{k_{C}+6-k}x^k}\, ,\quad
t_{23}(x)=\frac 1{x f(x)}{\sum\limits_{k=0}^{k_C+5}o^{CC}_{k_{C}+5-k}x^k}\, , \label{eq:pijtij}
\end{eqnarray}
with $s^{XY}_k$ and $o^{XY}_k$ being some positive constant integers. Using these relations, one can calculate $w_l$ and $q_l$, and by putting them, together with $x_l$, $u_l$, $v_l$ and $E_l$, into \eref{eq:otvoreneuv}, finally can evaluate $Z_{l}^O$, for any $l$.

In order to find asymptotic behavior of $Z_{l}^O$, we first notice that relations \eref{eq:wq} can be combined with the relation \eref{eq:xl} for $x_l$, so that one can establish recursion relations for $w_lx_l^2$ and $q_lx_l^3$, terms through which variables $w_l$ and $q_l$ appear in \eref{eq:otvoreneuv}. Using the facts that $x_l\to 0$, $f(x_l)\to p_{k_E}$, $u_l\to 0$, and $v_l\to\mathrm{const}$, when $l\to\infty$, it can be shown that
\begin{equation}
\left(
  \begin{array}{c}
    w_{l+1}x_{l+1}^2 \\
    q_{l+1}x_{l+1}^3 \\
  \end{array}
\right)\approx x_l^6\left[\left(
                    \begin{array}{cc}
                      a_{11} & a_{12} \\
                      a_{21} & a_{22} \\
                    \end{array}
                  \right)\left(
  \begin{array}{c}
    w_{l}x_{l}^2 \\
    q_{l}x_{l}^3 \\
  \end{array}
\right)+\left(
             \begin{array}{c}
               b_{13}v_l^2 \\
               0 \\
             \end{array}
           \right)\right]\, ,  \label{eq:wqasim}
           \end{equation}
where $a_{ij}$ and $b_{13}$ are some constants (see \ref{app:dh}). This relation implies that $w_lx_l^2$ and $q_lx_l^3$ tend to zero as $l\to\infty$, and consequently, comparing the terms containing $w_lx_l^2$ and $q_lx_l^3$ with those with $u_l$ and $v_l$ in \eref{eq:otvoreneuv}, one can conclude that
\[
\fl Z_{l+1}^O\approx E_l^{N_G}x_l^{N_G-k_C-3}\sum\limits_{k=0}^{k_C+1}\left(z_{k_C+1-k}^{AA}u_l^2+z_{k_C+1-k}^{AC}u_lv_l+
z_{k_C+1-k}^{CC}v_l^2\right)x_l^k\, .
\]
This means that for $l\gg 1$ number $Z_{l+1}^O$  behaves as
\[
Z_{l+1}^O \sim E_l^{N_G}x_l^{N_G-k_C-3}\, ,
\]
so that, using \eref{eq:zatvPomoc}, one obtains
\[
\frac{Z_{l+1}^O}{Z_{l+1}^C}\sim x_l^{-3}\sim \lambda^{-3^{l+1}}\, .
\]
Consequently, taking into account \eref{eq:scalingZatvorene}, it follows that overall number $Z_l^O$ of
open HWs  scales as
\begin{equation}
Z_l^O\sim \omega^{N_l}\mu_O^{N_l^\sigma}\, , \qquad
\mu_O=\lambda^{B}\, ,\qquad B=\frac{k_C+3}{N_G-3}\,4^{-\sigma}\, , \label{eq:scalingOtvorene}
\end{equation}
where, as in \eref{eq:sigma}, $\sigma=\ln 3/\ln N_G$,  and particular values of $\omega$, $\mu_O$ and
$\lambda$ are given in \tref{tab:rezhw}.

\section{Summary and Discussion \label{sec:discussion}}

In this paper we have analyzed asymptotic behavior of the
numbers of open and closed Hamiltonian walks on three-dimensional modified Sierpinski gasket family of fractals.
Numbers of extremely long HWs on these lattices can be generated by applying an exact recursive enumeration scheme, based on the fact that any HW on the $(l+1)$th step of the construction of the 3d MSG fractal, $G_{l+1}(b)$,  consists of $N_G=b(b+1)(b+2)/6$ HWs within its $N_G$ constitutive $G_l(b)$ structures, which can be of one of six possible types: $A, B, C, D, E$, and $H$ (\fref{fig:primerZatvorena} and \fref{fig:primerOtvorena}). Numbers of these HWs, $A_l$, $B_l$, $C_l$, $D_l$, $E_l$, and $H_l$,  fulfil closed set of recursive relations, \eref{eq:eb}, \eref{eq:ac} and \eref{eq:h}, which, due to the self-similarity of the lattices under study, do not depend on $l$, and therefore can be obtained by explicit enumeration and classification of HW conformations on the first step of the fractal construction, $G_1(b)$. As $b$ grows, number of HW conformations rapidly increases already on $G_1(b)$, so that we have managed to find explicit form of the recursive relations only up to $b=4$. However, we have shown that for any $b>2$ these recursive relations have some features that enable general analysis, leading to conclusion that overall numbers  $Z_l^C$ and $Z_l^O$ of closed and open HWs, respectively, on any 3d MSG fractal, scale with the number $N_l=4N_G^l$ of the lattice sites as
\[
Z_l^C\sim \omega^{N_l}\left[\lambda^{\frac{N_G+k_C}{N_G-3}4^{-\sigma}}\right]^{N_l^\sigma}\, , \quad
Z_l^O\sim \omega^{N_l}\left[\lambda^{\frac{k_C+3}{N_G-3}\,4^{-\sigma}}\right]^{N_l^\sigma}\, ,\]
with $\sigma$ given by formula \eref{eq:sigma}:
\[\sigma=\frac{\ln 3}{\ln\frac 16 b(b+1)(b+2)}=\frac{\ln 3}{\ln b}\frac 1{d_f}\,. \]
Constants $\omega$ and $\lambda$  can be obtained numerically, using relations
\[
\ln\omega=\lim_{l\to\infty}\frac{\ln E_{l}}{N_l}\, , \quad \ln\lambda=\lim_{l\to\infty}\frac{\ln(B_l/E_l)}{3^l}\, ,\]
and their particular values for $b=2,3$, and 4 are given in \tref{tab:rezhw}. Number $k_C=(b+1)(b+2)/2-8$ represents the maximal number of  $G_l(b)$ generators within the $G_{l+1}(b)$ they belong to, which are traversed by two-stranded $E$-type HW conformation ($E$-step) within any closed $HW$ on $G_{l+1}(b)$ (see equation \eref{eq:zatvorene} and \ref{ap:kb}). As one can see, HW conformations that traverse $G_l(b)$, and  recursion relations \eref{eq:eb} for the corresponding numbers $B_l$ and $E_l$, are sufficient for obtaining connectivity constant $\omega$ that governs the leading exponential term in the scaling forms for $Z_l^C$ and $Z_l^O$, as well as constant $\lambda$ which appears in their correction terms. Whereas the leading term $\omega^{N_l}$ is the same for both closed and open HWs, correction term in both cases has the same stretched exponential form $\mu^{N_l^\sigma}$, but with different values for $\mu$ and $\mu_O>\mu_C$, indicating that number of open HWs is larger than the number of closed HWs. One should notice here that the fact that scaling form obtained for open HWs is determined by the behavior of the numbers $B_l$ and $E_l$ only, means that contribution of one-- and two--leg HW conformations ($A$--, $C$--, $D$-- and $H$--type walks), {\em i.e.} HWs with their ends in the interior of $G_l(b)$ fractal structures, is not significant. Furthermore, in sections 2 and 3 it was elaborated that the main mathematical reason for obtaining scaling forms with the stretched exponential correction term is the fact that $\lim_{l\to\infty}B_l/E_l=0$,  meaning that entangled conformations (with large number of two-stranded parts) dominate over those in which HWs rarely return to already visited fractal generators.

Results of the study of HWs on 3d MSG fractals, presented in this paper, should be compared with the results recently obtained for other fractal lattices \cite{Elezovic}. For $n$-simplex fractal lattices with even values of $n$ the same asymptotic behavior of the HW numbers was found, $\omega^N\mu^{N^\sigma}$, but with $\sigma=1/d_f$, which differs from the formula derived here for $\sigma$. The existence of the term $\mu^{N^\sigma}$ was also related with the facts that: (1) entangled HWs prevail, and (2) HWs with "interior" terminating points do not contribute to the asymptotic behavior of HW numbers, as in the 3d MSG fractals case.  For odd $n$, stretched exponential terms in the scaling forms for HWs on $n$-simplex fractals do not exist, HWs with "interior" terminating points do affect the scaling forms for the numbers of open HWs, and entangled conformations do not dominate in this case. For Given-Mandelbrot and 2d modified Sierpinski gasket fractals (which are two-dimensional generalizations of the Sierpinski gasket and 3-simplex fractal, respectively) stretched exponential terms are also absent, whereas again, one-- and two--leg conformations are necessary for obtaining the scaling forms for open HWs, and, due to the specific topology of these lattices, only one-stranded HW conformations are possible. Therefore, results of the present study confirm the assumption that existence of the stretched exponential term is related with the issue of HW entanglement. However, the discrepancy between the formulas obtained for the exponent $\sigma$ for different fractal families, and, more generally, the question which properties of the underlying lattice and in what way determine $\sigma$, deserves further consideration. To this end we recall that stretched exponential terms were obtained for HWs on Manhattan \cite{Manhattan}, as well as for low-temperature SAWs on square \cite{Prellberg,SamoOwczarek,BennetWood,Baiesi} and cubic \cite{Grassberger} lattice. In this context, one should also mention the spiral SAWs on 2d regular lattices \cite{Guttman} and lattice animals on some hierarchical lattices. Whereas, to the best of our knowledge, physical interpretation of the stretched exponential term for the 2d spiral SAW models has never been proposed, its presence in the scaling forms obtained for lattice animals on hierarchical lattices  was directly connected with the existence of the sets of sites with different coordination numbers \cite{katarina}. Utilizing similar idea, in the case of low-temperature (collapsed) SAWs  it was explicitly explained in \cite{Owczarek} why term of the form $\mu^{N^\sigma}$, with $\sigma=(d-1)/d$ for regular lattices, should arise. However, as was elaborated in \cite{Stajic,Elezovic}, direct application of such approach on HWs on fractal lattices does not give satisfactory result. Yet, the scaling forms obtained in this paper can be expound in the spirit of the physical reasoning given in \cite{Owczarek}, as will be explained in the following paragraph.

We shall first focus on closed HWs on $G_l(b)$ with the maximal number of two-stranded $E$ conformations within the unit tetrahedrons. Such HWs represent maximally entangled closed (MEC) compact conformations and they accomplish the maximal possible connectedness between the generators $G_l(b)$ of all orders $l$. Now, from all the $N_l$ sites visited by such maximally entangled HW, observe those which belong to one-stranded conformations ($B$-steps) within the unit tetrahedrons, and which are directly connected only with the sites belonging to the same unit tetrahedron. For instance, vertices 9 and 10, or 15 and 16 in \fref{fig:primerZatvorena} represent examples of such sites. Since these sites are not directly connected with the other tetrahedrons of the lattice (via HW), they  are the maximally isolated sites within the maximally entangled closed HW conformation.  It is shown in \ref{app:diskusija} that number $N_l^{CI}$ of such sites is equal to
\begin{equation}
N_l^{CI}=\frac 12\frac{N_G-k_C-6}{N_G-3}N_l+2\frac{N_G+k_C}{N_G-3}4^{-\sigma}N_l^\sigma\, ,  \label{eq:nli}
\end{equation}
and one can easily  see that, using this expression, scaling relation for the numbers od closed HWs can be transformed into the following form
\begin{equation}
Z_l^C\sim {\omega'}^{N_l}{\sqrt{\lambda}}^{N_l^{CI}}\, ,
\quad \mathrm{with} \quad
\omega'=\omega/\lambda^{\frac{N_G-k_C-6}{4(N_G-3)}}\, .  \label{eq:konacnoZatv}
\end{equation}
In a similar way (see \ref{app:diskusija}), one can show that scaling relation for open HWs can be expressed as
\begin{equation}
Z_l^O\sim {\omega'}^{N_l}{\sqrt{\lambda}}^{N_l^{OI}}\, , \label{eq:konacnoOtv}
\end{equation}
where
\begin{equation}
N_l^{OI}=\frac 12\frac{N_G-k_C-6}{N_G-3}N_l+2\frac{k_C+3}{N_G-3}4^{-\sigma}N_l^\sigma \label{eq:nlio}\end{equation}
 is the number of maximally isolated sites on $G_l(b)$ visited by maximally entangled open HW (which is an open HW with the maximal number of $E$-steps on unit tetrahedrons and with both loose ends being of type $C$, on all levels $l$).
Comparing forms \eref{eq:konacnoZatv} and \eref{eq:konacnoOtv} with the scaling relations expected for HWs on homogeneous lattices, one can say that in this case terms ${\sqrt{\lambda}}^{N_l^{CI}}$ and ${\sqrt{\lambda}}^{N_l^{OI}}$ play the role of the stretched exponential correction term $\mu^{N^\sigma}$ in the case of homogeneous lattices. Indeed, for homogeneous lattices  $N^\sigma$, with $\sigma=(d-1)/d$, is proportional to the number of sites on the lattice boundary, which have smaller number of neighbors than the bulk sites, and in that sense they are similar to the maximally isolated sites within the maximally entangled HWs on 3d MSG lattices. Therefore, one might generally expect that scaling relation for the number of HWs on nonhomogeneous lattices instead of stretched exponential correction term has a term $\mu^{N_I}$, where $N_I$ is the number of conveniently defined "maximally isolated sites". Relation between $N_I$ and the overall number of lattice sites $N$ depends on the particular topology of the lattice under study, which is the reason why different scaling forms, as functions of $N$, are obtained.

Conclusion of the previous paragraph can be supported by performing a similar analysis in the case of HWs on previously studied fractal families \cite{Elezovic}. For instance, for $n$-simplex fractals maximally isolated sites can be defined in a similar way as for 3d MSG fractals, {\em i.e.} these are the sites which are directly connected by maximally entangled HW only with the sites within the same fractal generator they belong to. Here, maximally entangled HWs are those that contain maximal number of maximally-stranded HW parts. Then, it can be shown  \cite{nestampano} that $N_l^{CI}=\frac n2 N_l^{\sigma}$ and $N_l^{OI}=\left(\frac n2-1\right)N_l^{\sigma}$, where $\sigma=\ln 2/\ln n$. Since overall numbers of closed and open HWs on $n$-simplex fractal with even $n$ asymptotically behave as $Z_l^C\sim \omega^{N_l}\lambda_B^{\frac n2 N_l^{\sigma}}$ (see (5.8) in \cite{Elezovic}) and $Z_l^O\sim \omega^{N_l}\lambda_B^{\frac{n-2}2 N_l^{\sigma}}$ (formula (5.14) in \cite{Elezovic}), respectively, they immediately fit into the scaling relation $\omega'^N\mu^{N_I}$, where $\omega'=\omega$ and $\mu=\lambda_B$ for both types of HWs. For odd values of $n$, however, it can be shown \cite{nestampano} that number $N_I$ of maximally isolated sites, for both open and closed HWs, is proportional to $N_l$, so that any term of the form $\mu^{N_I}$ can only contribute to the leading term $\omega^N$.  This certainly is in accord with the scaling forms
found in \cite{Elezovic} for odd $n$, which do not have stretched exponential corrections.

To conclude, we may say that presented exact study of Hamiltonian walks on modified three-dimensional Sierpinski gasket fractals supports the idea that entangled conformations are of the most physical importance for the behavior of compact polymers in inhomogeneous media. The introduced concept of "maximally isolated sites" proved to be useful in physical interpretation of different scaling forms obtained for Hamiltonian walks on all so far studied fractals. Whether such concept can be applied on homogeneous and other nonhomogeneous lattices will be the matter of our further investigations.

\ack

This paper has been done within the Project No OI 141020B, funded by the Serbian Ministry of Science and Environmental
Protection.

\appendix

\section{Maximal number of $E$-steps within HW configurations\label{ap:kb}}

In this Appendix we prove that numbers $k_C$, $k_B$ and $k_E$ are given by formulae \eref{eq:kc} and \eref{eq:kbke}. The number $k_C$ is the maximum number of $E$-steps within the closed HW configuration (see \eref{eq:zatvorene}), whereas the numbers $k_B$ and $k_E$ are related to the maximum number of $E$-steps within the HW configurations of $B$ and $E$-type (see \eref{eq:eb}), {\em i.e.} the conformations that traverse the generator of order $l+1$ once or two times, respectively. By ,,$E$-step'' here we imply any HW conformation consisting of two mutually avoiding strands, both traversing the generator of order $l$ (see \fref{fig:primerZatvorena}). In a similar way, ,,$B$-step'' is any HW conformation that consists of one strand traversing $G_l(b)$.

\begin{figure} \hskip2cm
\includegraphics[height=.4\textheight]{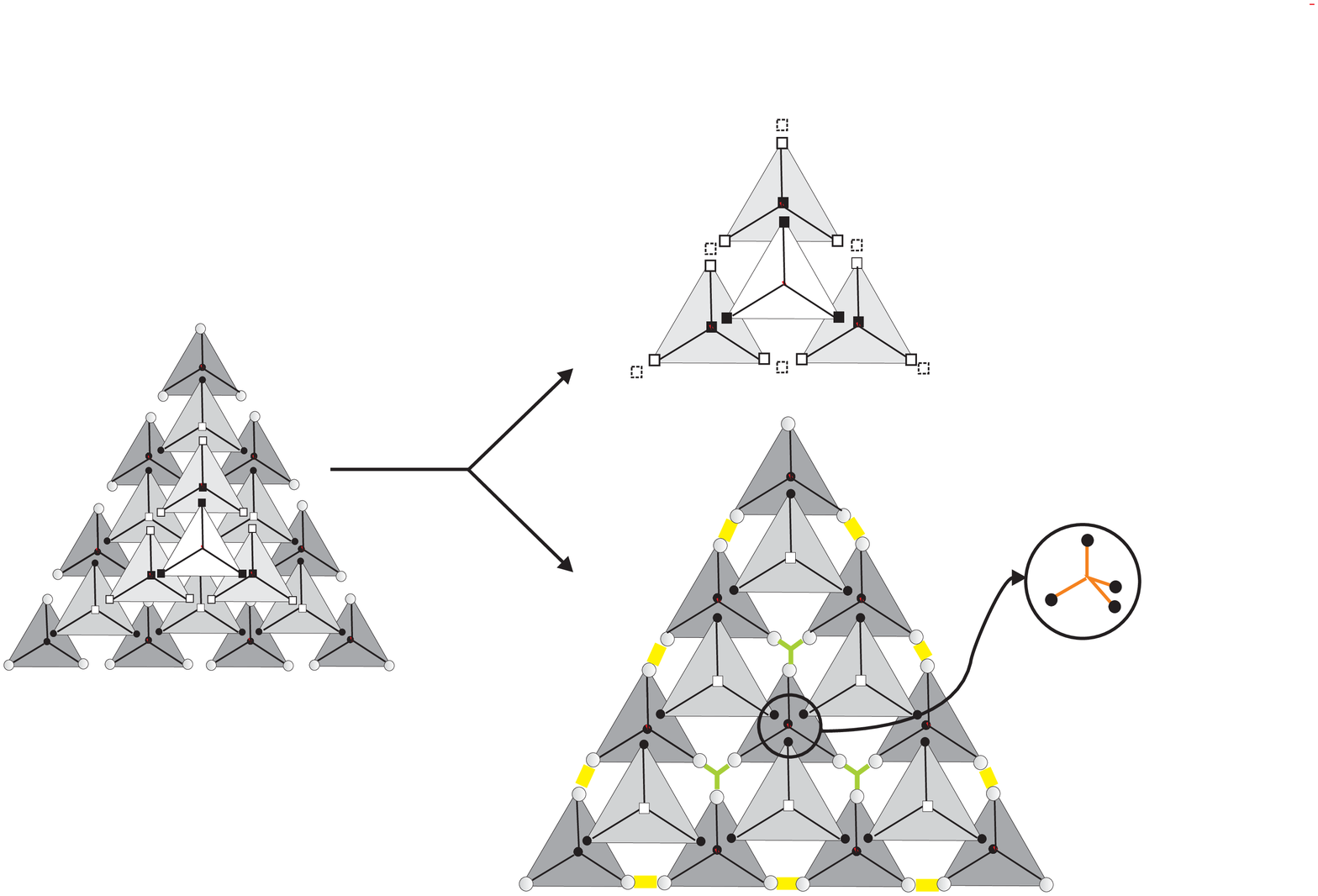} \caption{On the left-hand side of this picture generator $G_{l+1}(b)$ of order $l+1$, for $b=4$ 3d MSG, is presented, as it is seen from above. Gray-shaded small tetrahedrons represent generators $G_l(4)$ of order $l$, lower layers being darker. Vertices lying in the same horizontal plane are indicated with the same symbol, being also the same for two adjacent planes containing vertices that belong to different generators. On the right-hand side of the picture horizontal layers of tetrahedrons are split and slightly magnified in order that all vertices, as well as some junctions can be seen. In particular, twofold junctions in the lowest layer of vertices are indicated by yellow lines, whereas junctions joining three vertices are given in green color. There is only one four-fold junction -- the corresponding four connected vertices are encircled, and in the magnified circle these vertices are presented as seen in three dimensions, with the junction indicated in orange color. One can check that overall number of junctions is $N_J=31$ (three twofold junctions for each of six $G_{l+1}(4)$ edges, three three-fold junctions for each of four $G_{l+1}(4)$ faces, and one four-fold interior junction), which certainly is in accord with formula \eref{eq:brojveza}.}
\label{fig:b4primer}
\end{figure}

According to our definition of the modified Sierpinski gasket fractals, generators $G_l(b)$ within the generator $G_{l+1}(b)$ are connected via infinitesimal junctions. Each of these junctions connects two, three or four neighboring $G_l(b)$ generators. ,,Twofold'' junction connects vertices of neighboring $G_l(b)$, both lying in the same edge of $G_{l+1}(b)$ (for instance, vertices 3 and 4 in \fref{fig:primerZatvorena} are joined by such junction). Vertices of neighboring $G_l(b)$ that lie inside the faces of $G_{l+1}(b)$ (such as sites 5,6 and 19 in \fref{fig:primerZatvorena}) are connected by three-folded junction, whereas four-folded junction occur in the interior of $G_{l+1}(b)$, and can exist only for $b\geq 4$. In \fref{fig:b4primer} we explicitly indicate some of the junctions connecting the vertices of the generators $G_l(4)$, within the generator $G_{l+1}(4)$.
It is not difficult to show that the number of junctions which connect $N_G$
generators $G_l(b)$ into the generator $G_{l+1}(b)$ is equal to
\begin{equation}
N_J=\frac 16 (b+1)(b+2)(b+3)-4\, .
\label{eq:brojveza}
\end{equation}
From the restraint that each site of the lattice has to be visited exactly once,
follows that each junction ({\em i.e.} its middle point) can be traversed at most once. Now, suppose that closed HW
conformation within $G_{l+1}(b)$ consists of $\alpha$ $B$-steps and $\beta$ $E$-steps. Each $G_l(b)$ is traversed by
HW, which implies that
\begin{equation} \alpha+\beta=N_G=\frac 16 b(b+1)(b+2)\, .  \label{eq:veze1}
\end{equation}
Each
$B$--step uses two junctions and each $E$--step uses four junctions. However, since each junction connects two $G_l(b)$
generators, the overall number of visited junctions is equal to $\alpha+2\beta$. In addition, steps through corner
$G_{l}(b)$ generators for the closed HW have to be of the $B$-type (see \fref{fig:primerZatvorena}), which means that
one of the three junctions for each of four corner generators is certainly not used.  This implies that inequality
\begin{equation}
\alpha+2\beta\leq N_J-4=\frac 16 (b+1)(b+2)(b+3)-8 \label{eq:veze2}
\end{equation}
is satisfied. Then, from \eref{eq:veze1} and \eref{eq:veze2} directly follows that
\begin{equation}
\beta\leq \frac 12 (b+1)(b+2)-8\, ,
\end{equation}
 {\em i.e.} the maximum number of $E$-steps within the closed HW conformation is indeed equal to
\begin{equation}
\beta_{\mathrm{max}}=k_C=\frac 12 (b+1)(b+2)-8\, . \label{eq:betamaxkc}
\end{equation}
\begin{figure}
\hskip1cm \includegraphics[height=.3\textheight]{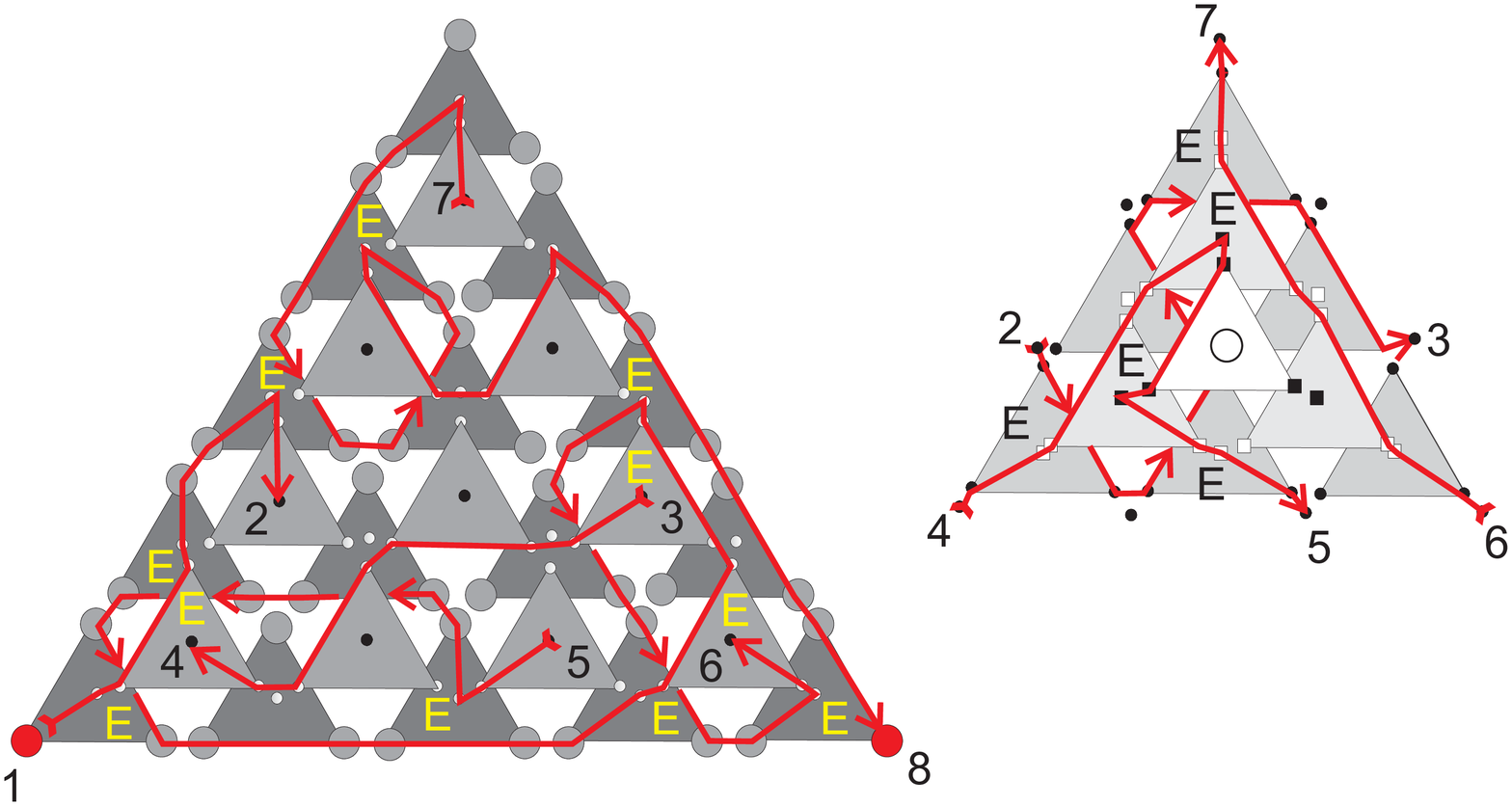}
\caption{Generator $G_{l+1}(5)$, viewed from above, with the upper three layers of constitutive generators $G_l(5)$  (gray-shaded tetrahedrons-triangles) moved to the right-hand side of the picture, for the sake of better visuality. Red oriented line represents a $B$-type HW configuration, which contains 16 $E$-steps, {\em i.e.} the maximum possible number of them, given by formula \eref{eq:betamaxB}. Remaining 19 steps are of the $B$-type, and they are depicted as straight lines connecting only two vertices of the traversed $G_l(5)$ generator, but it is implied that all sites within it are visited. The numbers in the figure denote points where HW leaves or enters the lower part of $G_{l+1}(5)$.} \label{fig:b5primerB}
\end{figure}

Next, we consider a $B$--type HW configuration on $G_{l+1}(b)$ generator, with $\alpha$ $B$--steps, and $\beta$ $E$--steps. Steps through two corner $G_l(b)$ generators, at which HW starts or terminates, can be of either $B$-- or $E$--type. However, both remaining two corner $G_l(b)$ generators must be traversed by $B$--step (see \fref{fig:b5primerB}), meaning that two of $N_J$ junctions are never used. On the other hand, each $B$--step uses two junctions, unless it is the first or the last step of the HW configuration, in which case it utilizes only one junction (since the corner vertex does not belong to any junction). Similarly, $E$--step which traverses interior $G_l(b)$ generators uses four junctions, whereas possible $E$--step through either the first or the last traversed $G_l(b)$ uses only three junctions. Suppose that both the first and the last visited corner generators contain $E$-step. Then, these two steps together use six junctions, and the remaining $(\beta-2)$ ,,interior'' $E$--steps use $4(\beta-2)$, whereas $\alpha$ $B$--steps use $2\alpha$ junctions. Since every junction can be traversed only once, and two consecutive steps share junctions, it follows that altogether $[6+4(\beta-2)+2\alpha]/2$ junctions are visited by such HW conformation. This implies the following inequality
\begin{equation} \frac 12[6+4(\beta-2)+2\alpha]\leq N_J-2\, ,
\end{equation}
which, together with the relation $\alpha+\beta=N_G$, gives
\begin{equation} \beta\leq N_J-N_G-1=\frac
12(b+1)(b+2)-5=\beta_{\mathrm{max}}=N_G-k_B\, .  \label{eq:betamaxB}
\end{equation}
In the remaining two possible situations, when the first and the last $G_l(b)$ are traversed by either both $B$--steps, or one of them by $B$--step and the other by $E$--step, in a similar way it can be shown that inequality \eref{eq:betamaxB} holds as well.
\begin{figure} \hskip1cm \includegraphics[height=.3\textheight]{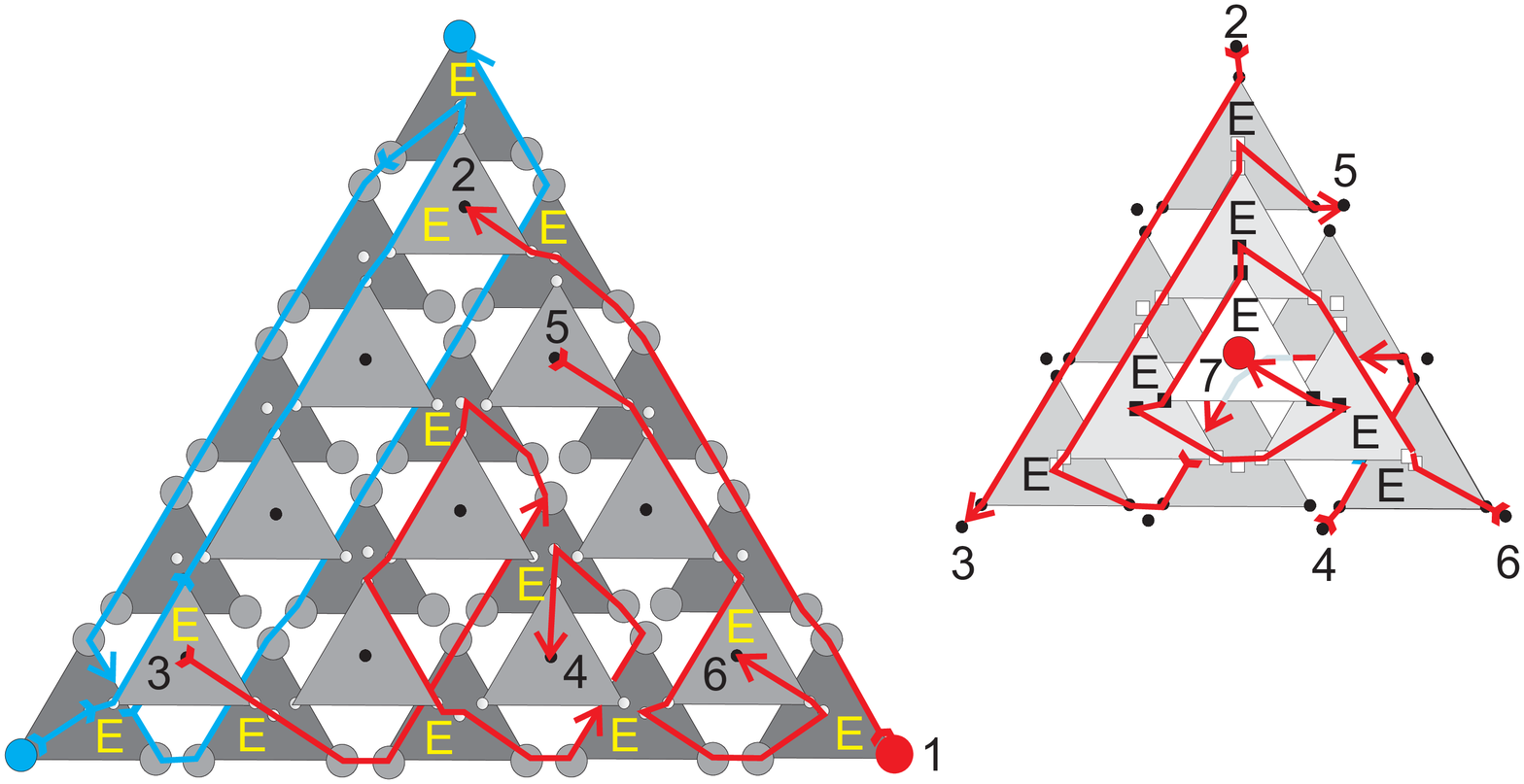} \caption{Example of $E$--type HW
conformation on $G_{l+1}(5)$ with 19 $E$--steps, which is the maximum possible number $\beta_{\mathrm{max}}$ of them,
given by formula \eref{eq:betamaxE}. The two strands of the conformation are depicted as oriented red and blue lines.
The red strand starts at vertex labeled with number 1, and terminates at vertex 7, whereas the numbers 2 to 6 mark the
sites at which the strand leaves or enters the second layer of $G_{l}(5)$ generators.} \label{fig:b5primerE}
\end{figure}

As one can see in \fref{fig:b5primerE} an $E$--type HW conformation within $G_{l+1}(b)$ can use all three
junctions of any corner $G_{l}(b)$, which happens when such $G_{l}(b)$ is traversed by $E$-step. This also means that
all $N_J$ junctions within $G_{l+1}(b)$ can be utilized. Suppose that all four corner $G_{l}(b)$ generators are
traversed by $E$--steps -- altogether 12 junctions are used by these four $E$-steps. If the whole HW conformation
contains $\beta$ $E$--steps, the remaining $(\beta-4)$ interior $E$-steps take $4(\beta-4)$ junctions, whereas to all
$\alpha$ $B$--steps $2\alpha$ junctions correspond. Every used junction connects two steps, therefore in this case
inequality
\begin{equation}
\frac 12[12+4(\beta-4)+2\alpha]\leq N_J\, , \end{equation} holds, which, again with
$\alpha+\beta=N_G$, leads to \begin{equation} \beta\leq N_J-N_G+2=\beta_{\mathrm{max}}=\frac 12(b+1)(b+2)-2=N_G-k_E\, .
\label{eq:betamaxE}
\end{equation}
In a similar manner one can show that in the remaining four possibilities: one, two,
three or four $B$--steps through the corner $G_l(b)$ generators, inequality \eref{eq:betamaxE} is valid again.

Note that formulae \eref{eq:betamaxkc}, \eref{eq:betamaxB}  and \eref{eq:betamaxE} are not correct in the $b=2$ case.
In this case all four constitutive $G_l(b=2)$ generators of the $G_{l+1}(b=2)$ generator
are at its corners, and each of them is connected with each of the other three by two-folded junction. Since two corner generators have a junction in common, the number  of usable junctions in relations (A.3) and (A.6) is $N_J-2$ and $N_J-1$ instead of  $N_J-4$ and  $N_J-2$, respectively. If we represent each $G_l(2)$ by a point (vertex) then $G_{l+1}(2)$  represents a complete graph of four points. Since there can be neither subgraph with three points of degree three and one point of degree one, nor subgraph with two points of degree three and two of degree one,  terms in relation \eref{eq:rrb2} with $E^3$ and $E^2$ are not possible.

\section{Asymptotical behavior of the parameter $x_l=B_l/E_l$\label{app:xasimptotika}}

As it was stressed in section~\ref{sec:zatvorene} asymptotical behavior of the parameter $x_l=B_l/E_l$ is of the greatest importance for obtaining the scaling form of HWs. Here we want to prove that assumption $p_{k_E+k}>m_{k_E+k+1}$, for $2\leq k\leq k_C+5$, leads to the conclusion that $x_l\to 0$, when $l\to\infty$. Numbers $x_l$ satisfy the exact relation \eref{eq:xl}, from which follows that
\[
\fl x_{l+1}-x_l=x_l\frac{-p_{k_E}-p_{k_E+1}x_l-p_{k_E+k_C+6}x_l^{k_C+6}-
\sum\limits_{k=2}^{k_C+5}(p_{k_E+k}-m_{k_E+k+1})\, x_l^k}{\sum\limits_{k=0}^{k_C+6}p_{k_E+k}\, x_l^k}<0\, ,\]
where we used the relation $k_E=k_B-3$, obtained in the previous Appendix, the fact that all coefficients $p_k$ and $m_k$ are positive, as well as the proposed inequality. This means that sequence of positive numbers $x_l$ is monotonically decreasing, and since $x_0=2$, it follows that this sequence has non-negative limiting value $c=\lim_{l\to\infty}x_l$. If $c>0$, again from \eref{eq:xl}, when $l\to\infty$, follows equation
\[c=c^3\frac{\sum\limits_{k=0}^{k_C+3}m_{k_B+k}\, c^k}{\sum\limits_{k=0}^{k_C+6}p_{k_E+k}\, c^k}\, ,
\]
which, after dividing by $c$, can be transformed to
\[
0=p_{k_E}+p_{k_E+1}c+p_{k_E+k_C+6}c^{k_C+6}+
\sum\limits_{k=2}^{k_C+5}(p_{k_E+k}-m_{k_E+k+1})c^k\, .\]
However, the latter equation cannot be satisfied, since its right-hand side has the form of polynomial in $c$, with all its coefficients being positive numbers. Consequently, one concludes that $c=0$.

\section{Recursion relations for the numbers $A_l$, $C_l$ and $Z_{l}^O$ \label{app:ac}}

In this Appendix we first show that polynomials $R_{ij}$, appearing in recursion relations \eref{eq:ac} for numbers
$A_l$ and $C_l$ of one-leg HW conformations,  have the form \eref{eq:r}, with  $k_{ij}$ given by \eref{eq:rgranice}.
It is obvious that dangling end-step  of one-leg configuration within $G_{l+1}(b)$ can be either of $A$-- or $C$--type. The remaining part of the HW traverses the remaining $N_G-1$ generators $G_l(b)$, and consists of $\alpha$ $B$--steps and $\beta$ $E$--steps, so that equation
\begin{equation}
\alpha+\beta=N_G-1\, \label{eq:ng-1}
\end{equation}
is satisfied.

Consider an $A$--type HW configuration on generator  $G_{l+1}(b)$. One of its end-steps is fixed at one of the four
corner generators $G_l(b)$ and it can be either $B$--, $E$--, or $C$--step, respectively using one, three, or two
junctions. Two of the remaining three corner generators are traversed by $B$--steps, each of them using two junctions of the corresponding $G_l(b)$, whereas its third junction cannot be traversed. The last corner generator is either traversed by a $B$--step, or it contains dangling $A$-- or $C$--step, and such steps use two, one, or three junctions, respectively. If a corner generator contains $A$--step, only one of its corresponding three junctions can be traversed. Now, suppose that ,,corner'' steps are $EBBB$, whereas the dangling $A$-step, using one junction, is in the interior generator. Such HW traverses $(3+2\alpha+4(\beta-1)+1)/2=\alpha+2\beta$ junctions, and, since corner $B$-steps cannot use altogether three of all $N_J$ junctions, it follows that
\begin{equation}
\alpha+2\beta\leq N_J-3\, , \label{eq:nejednakost}
\end{equation}
which with \eref{eq:ng-1} gives
\begin{equation}
\beta\leq N_J-N_G-2=\frac 12(b+1)(b+2)-6=k_C+2\, .
\end{equation}
In a similar manner one can analyze all the other possibilities and obtain  corresponding maximum value of $\beta$. From the \tref{tab:Akonfiguracije}, in which all of these possibilities, together with the corresponding numbers of used and
usable junctions, as well as $\beta_ {\mathrm{max}}$ are given, one can see  that $k_{11}=k_{12}=k_C+2$.
\begin{table}
\caption{\label{tab:Akonfiguracije}All possible situations for $A$--type HW configuration on $G_{l+1}(b)$,
depending on the types of its dangling end-step, steps traversing corner generators $G_l(b)$ and end-step fixed in one
of the corners. For each situation the corresponding numbers of passed and usable junctions, as well as the maximum
possible value of number $\beta$ of $E$--steps, are given.}
\raggedleft \footnotesize\rm \begin{tabular}{@{}ccccccc} \br
dangling end-step &,,corner''  &,,corner'' &number of  &number of  \\
 and its position
&passing-steps & fixed-step& used junctions &usable junctions&$\beta_{max}$ \\
 \mr
$A$ - interior $G_l(b)$&$BBB$&$E$&$\alpha+2\beta$&$N_J-3$&\fbox{$N_J-N_G-2$}\\
 $A$ - interior$G_l(b)$&$BBB$&$B$&$\alpha+2\beta$&$N_J-5$&$N_J-N_G-4$\\
 $A$ - corner $G_l(b)$&$BB$&$E$&$\alpha+2\beta$&$N_J-4$&$N_J-N_G-3$\\
 $A$ - corner $G_l(b)$&$BB$&$B$&$\alpha+2\beta$&$N_J-6$&$N_J-N_G-5$\\
 \mr
  $C$ - interior $G_l(b)$&$BBB$&$E$&$1+\alpha+2\beta$&$N_J-3$&$N_J-N_G-3$\\
  $C$ - interior $G_l(b)$&$BBB$&$B$&$1+\alpha+2\beta$&$N_J-5$&$N_J-N_G-5$\\
  $C$ - corner $G_l(b)$&$BB$&$E$&$1+\alpha+2\beta$&$N_J-2$&\fbox{$N_J-N_G-2$}\\
  $C$ - corner $G_l(b)$&$BB$&$B$&$1+\alpha+2\beta$&$N_J-4$&$N_J-N_G-4$\\
  $C$ - corner $G_l(b)$&$BBB$&$C$&$1+\alpha+2\beta$&$N_J-4$&$N_J-N_G-4$\\
 \br
\end{tabular}
\end{table}

$C$--type HW configuration on $G_{l+1}(b)$ consists of two strands: one of $B$--type, {\em i.e.} with two ends fixed at
two corners of the $G_{l+1}(b)$, and the other of $A$--type, with one end fixed at the third corner, whereas its second
end is free, {\em i.e.} it can be anywhere within the $G_{l+1}(b)$, in some of interior generators $G_l(b)$, as well as
in any of the four corner-generators $G_l(b)$. The three steps containing fixed ends can be of either $B$-- or
$E$--type, whereas the corner-generator, which does not contain the fixed-end, is either traversed by a $B$--step or it
contains the dangling $A$-- or $C$--step. The other relevant details of possible situations, regarding the type and the
position of the step containing the dangling end, are listed in \tref{tab:Ckonfiguracije}. One can see that HW
configurations with $A$ dangling end-step and maximal number of $E$--steps are those with the dangling end-step in the
interior $G_l(b)$, whereas the three steps with ends fixed at corner-generators are of $E$--type, and the fourth
corner-generator is traversed by a $B$--step, and thus conclude that $k_{21}=N_J-N_G+1=k_C+5$. Configurations with $C$
dangling end and maximal number of $E$--steps have also all three fixed-steps of $E$--type, while the dangling end is in the fourth corner-generator, therefore $k_{22}=k_C+5$.
\begin{table}
\caption{\label{tab:Ckonfiguracije} Description of possible $C$--type HW configurations on $G_{l+1}(b)$,
depending on the types of dangling end-step, and the steps within corner-generators $G_l(b)$. For each situation the
corresponding numbers of passed and usable junctions, as well as the maximum
possible value of number $\beta$ of $E$--steps, are given.}
\raggedleft \footnotesize\rm \begin{tabular}{@{}ccccccc} \br
dangling end-step &,,corner''  &,,corner'' &number of  &number of  \\
 and its position
&fixed-steps & passing-step& used junctions &usable junctions&$\beta_{max}$ \\
 \mr
$A$ - interior $G_l(b)$&$BBB$&$B$&$\alpha+2\beta-1$&$N_J-7$&$N_J-N_G-5$\\
$A$ - interior $G_l(b)$&$BBE$&$B$&$\alpha+2\beta-1$&$N_J-5$&$N_J-N_G-3$\\
$A$ - interior $G_l(b)$&$BEE$&$B$&$\alpha+2\beta-1$&$N_J-3$&$N_J-N_G-1$\\
$A$ - interior $G_l(b)$&$EEE$&$B$&$\alpha+2\beta-1$&$N_J-1$&\fbox{$N_J-N_G+1$}\\
$A$ - corner $G_l(b)$&$BBB$&$A$&$\alpha+2\beta-1$&$N_J-8$&$N_J-N_G-6$\\
$A$ - corner $G_l(b)$&$BBE$&$A$&$\alpha+2\beta-1$&$N_J-6$&$N_J-N_G-4$\\
$A$ - corner $G_l(b)$&$BEE$&$A$&$\alpha+2\beta-1$&$N_J-4$&$N_J-N_G-2$\\
$A$ - corner $G_l(b)$&$EEE$&$A$&$\alpha+2\beta-1$&$N_J-2$&$N_J-N_G$\\
 \mr
$C$ - interior $G_l(b)$&$BBB$&$B$&$\alpha+2\beta$&$N_J-7$&$N_J-N_G-6$\\
$C$ - interior $G_l(b)$&$BBE$&$B$&$\alpha+2\beta$&$N_J-5$&$N_J-N_G-4$\\
$C$ - interior $G_l(b)$&$BEE$&$B$&$\alpha+2\beta$&$N_J-3$&$N_J-N_G-2$\\
$C$ - interior $G_l(b)$&$EEE$&$B$&$\alpha+2\beta$&$N_J-1$&$N_J-N_G$\\
$C$ - corner $G_l(b)$&$BBB$&$C$&$\alpha+2\beta$&$N_J-6$&$N_J-N_G-5$\\
$C$ - corner $G_l(b)$&$BBE$&$C$&$\alpha+2\beta$&$N_J-4$&$N_J-N_G-3$\\
$C$ - corner $G_l(b)$&$BEE$&$C$&$\alpha+2\beta$&$N_J-2$&$N_J-N_G-1$\\
$C$ - corner $G_l(b)$&$EEE$&$C$&$\alpha+2\beta$&$N_J$&\fbox{$N_J-N_G+1$}\\
$C$ - corner $G_l(b)$&$BBC$&$B$&$\alpha+2\beta$&$N_J-6$&$N_J-N_G-5$\\
$C$ - corner $G_l(b)$&$BEC$&$B$&$\alpha+2\beta$&$N_J-4$&$N_J-N_G-3$\\
$C$ - corner $G_l(b)$&$EEC$&$B$&$\alpha+2\beta$&$N_J-2$&$N_J-N_G-1$\\
 \br
\end{tabular}
\end{table}

In the remaining part of this Appendix we show that polynomials $F$ and $S$, appearing in the recursion relation
\eref{eq:otvorene} for the overall number $Z_{l+1}^O$ of open HWs on $G_{l+1}(b)$, have the form given by \eref{eq:fxy}. Polynomials $F_{AA}$, $F_{AC}$ and $F_{CC}$ correspond to HWs whose two step-ends are two $A$--steps, one $A$-- end one
$C$--step, and two $C$--steps, respectively. Since the whole HW configuration has $N_G$ steps,  its ,,interior'' part
has $N_G-2$ steps which are either of $B$-- or $E$--type, implying that   $F_{XY}$ are homogeneous polynomials in $B_l$
and $E_l$ of power $N_G-2$. In \tref{tab:fxy} we list all possible arrangements of steps within open HW configuration
with two one-leg end-steps, together with the corresponding numbers of used and usable junctions. Using these data one
can recognize arrangements with maximum possible number of $E$--steps, and conclude that this number is equal to
$N_J-N_G-3=k_C+1$.
\begin{table}
\caption{\label{tab:fxy} Possible steps arrangements of open HW configuration with one-leg ends, together with the
corresponding numbers of used and usable junctions. The last column represents the maximum number of $E$--steps for each arrangement, obtained by the fact that number of used junctions cannot be larger than the number of usable junctions.}
\raggedleft \footnotesize\rm \begin{tabular}{@{}ccccc} \br
steps through  &steps through   &number of used  &number of  usable &$\beta_{\mathrm{max}}$  \\
corner-generators & interior generators  & junctions & junctions&  \\
$AABB$&$(\alpha-2)$ $B$, $\beta$ $E$&$1+\alpha+2\beta$&$N_J-6$&$N_J-N_G-5$\\
$ABBB$&$(\alpha-3)$ $B$, $\beta$ $E$, 1 $A$&$1+\alpha+2\beta$&$N_J-5$&$N_J-N_G-4$\\
$BBBB$&$(\alpha-4)$ $B$, $\beta$ $E$, 2 $A$&$1+\alpha+2\beta$&$N_J-4$&\fbox{$N_J-N_G-3$}\\
  \mr
$ACBB$&$(\alpha-2)$ $B$, $\beta$ $E$&$2+\alpha+2\beta$&$N_J-4$&$N_J-N_G-4$\\
$ABBB$&$(\alpha-3)$ $B$, $\beta$ $E$, 1 $C$&$2+\alpha+2\beta$&$N_J-5$&$N_J-N_G-5$\\
$CBBB$&$(\alpha-3)$ $B$, $\beta$ $E$, 1 $A$&$2+\alpha+2\beta$&$N_J-3$&\fbox{$N_J-N_G-3$}\\
$BBBB$&$(\alpha-4)$ $B$, $\beta$ $E$, 1 $A$, 1 $C$&$2+\alpha+2\beta$&$N_J-4$&$N_J-N_G-4$\\
 \mr
$CCBB$&$(\alpha-2)$ $B$, $\beta$ $E$&$3+\alpha+2\beta$&$N_J-2$&\fbox{$N_J-N_G-3$}\\
$CBBB$&$(\alpha-3)$ $B$, $\beta$ $E$, 1 $C$&$3+\alpha+2\beta$&$N_J-3$&$N_J-N_G-4$\\
$BBBB$&$(\alpha-4)$ $B$, $\beta$ $E$, 2 $C$&$3+\alpha+2\beta$&$N_J-4$&$N_J-N_G-5$\\
 \br
\end{tabular}
\end{table}

If an open HW has both its ends in the same $G_l(b)$, then this corresponds to $D$-- or $H$--step within it, whereas
remaining part of that HW consists of $\alpha$ $B$-steps and $\beta$ $E$--steps such that $\alpha+\beta=N_G-1$. This
means that $S_D$ and $S_H$ are homogeneous polynomials in $B_l$ and $E_l$ of power $N_G-1$. An $H$--step can exist only
in interior $G_l(b)$, in which case all steps through corner-generators are of $B$-type, and four junctions can
never be utilized. Since such HW takes $2+\alpha+2\beta$ junctions, it follows that $\beta\leq N_J-N_G-5=k_C-1$.
Finally, $D$--step can exist in any $G_l(b)$, either corner or interior one, but in both cases the number of used
junctions is $1+\alpha+2\beta$, and four junctions corresponding to corner-generators cannot be taken by such HW. Thus,
one obtains that maximum number of $E$--steps is $\beta_{\mathrm{max}}=N_J-N_G-4=k_C$.

\section{Two-leg HW configurations \label{app:dh}}

Here we give some details of derivation of the conclusion that numbers of two-leg HW configurations are not needed for obtaining the scaling form of the overall number of open HWs.

First we notice that recursion relation for the two-leg HW configuration of the type $D$, given in \eref{eq:h}, follows from the fact that dangling ends of the two strands can reside either (1) in the same generator $G_l(b)$, thus forming $D$-- or $H$--step within it, or (2) in two different generators $G_l(b)$, thus forming two one-leg steps, either of $A$-- or $C$--type. It is not difficult to see that case (1) can be obtained by cutting a step of some $B$--type configuration. Therefore, each $B$--type configuration with $k$ $B$--steps and $(N_G-k)$ $E$--steps gives rise to $k$ $D$--type configurations with $(k-1)$ $B$--steps, $(N_G-k)$ $E$--steps and one $D$--step, which are obtained by cutting a $B$--step, whereas by cutting its $E$--steps (which can be done in two ways for each $E$--step) one can obtain $2(N_G-k)$ different $D$--type configurations with $k$ $B$--steps, $(N_G-k-1)$ $E$--steps and one $H$--step. In a similar way, by cutting an $E$--type HW configuration, one can obtain an $H$--type configuration. From these observations straightforwardly follows that
\begin{eqnarray}
\fl d_D(B,E)&=&\sum_{k=k_B}^{N_G}km_kB^{k-1}E^{N_G-k}\, , \quad
d_H(B,E)=2\sum_{k=k_B}^{N_G}(N_G-k)m_kB^{k}E^{N_G-k-1}\, ,\nonumber \\
\fl h_D(B,E)&=&\frac 12\sum_{k=k_E}^{N_G}kp_kB^{k-1}E^{N_G-k}\, , \quad
h_H(B,E)=\sum_{k=k_E}^{N_G}(N_G-k)p_kB^{k}E^{N_G-k-1}\, .  \label{eq:dDHhDH}
\end{eqnarray}
Applying a reasoning similar to one used in the previous Appendix, one can obtain that polynomials $d_{XY}$ and $h_{XY}$, corresponding to two-leg configurations with dangling ends in different generators, have the following forms
\begin{equation}
\fl d_{XY}(B,E)=\sum\limits_{k=0}^{k_C+4}s_k^{XY}B^{N_G-2-k}E^{k}\, ,\qquad
h_{XY}(B,E)=\sum\limits_{k=0}^{k_{XY}}o_k^{XY}B^{N_G-2-k}E^k\, ,  \label{eq:dXYhXY}
\end{equation}
with $s^{XY}_{k}$ and $o_k^{XY}$ being constant positive integers, $XY=AA,AC,CC$, and
\[
k_{AA}=k_C+7=k_{AC}+1=k_{CC}+2\, .
\]

Next, recursion relations \eref{eq:wq} are obtained by dividing  \eref{eq:h} with the recursion relation for $E_l$, given in \eref{eq:eb}. Therefore, coefficients $p_{ij}$ are defined as
\begin{eqnarray}
p_{11}(x_l)&=& \frac{d_D(B_l,E_l)E_l}{E_{l+1}(B_l,E_l)}\, , \qquad p_{12}(x_l)=
\frac{d_H(B_l,E_l)E_l}{E_{l+1}(B_l,E_l)}\, ,\nonumber \\
p_{21}(x_l)&=& \frac{h_D(B_l,E_l)E_l}{E_{l+1}(B_l,E_l)}\, , \qquad p_{22}(x_l)=
\frac{h_H(B_l,E_l)E_l}{E_{l+1}(B_l,E_l)}\, , \nonumber
\end{eqnarray}
and $t_{ji}$  as
\[
t_{1i}(x_l)=\frac{d_{XY}(B_l,E_l)E_l^2}{E_{l+1}(B_l,E_l)}\, , \qquad
t_{2i}(x_l)=\frac{h_{XY}(B_l,E_l)E_l^2}{E_{l+1}(B_l,E_l)}\, ,
\]
where $i=1,2,3$ corresponds to $XY=AA, AC, CC$, respectively. Then, from \eref{eq:dDHhDH} and \eref{eq:dXYhXY} relations \eref{eq:pijtij} directly follow.

For $l\gg 1$, $x_l$ tends to 0, so that from \eref{eq:pijtij} one obtains
\begin{eqnarray}
\fl p_{11}(x_l)\approx k_B\frac{m_{k_B}}{p_{k_E}}x_l^2\, , \qquad p_{12}(x_l)\approx 2(k_C+3)\frac{m_{k_B}}{p_{k_E}}x_l^3\, , \qquad
p_{21}(x_l)\approx\frac{k_E}{2}\frac 1{x_l}\, , \nonumber\\ \fl p_{22}(x_l)\approx k_C+6\, , \qquad
t_{1i}(x_l)\approx\frac{s_{k_C+4}^{XY}}{p_{k_E}}\, , \qquad t_{21}(x_l)\approx\frac{o_{k_C+7}^{AA}}{p_{k_E}}\frac{1}{x_l^3}\, ,\nonumber\\
\fl t_{22}(x_l)\approx\frac{o_{k_C+6}^{AC}}{p_{k_E}}\frac 1{x_l^2}\, , \qquad
t_{23}(x_l)\approx\frac{o_{k_C+5}^{CC}}{p_{k_E}}\frac 1{x_l}\, . \nonumber
\end{eqnarray}
Finally, using these approximate forms while multiplying \eref{eq:wq} with \eref{eq:xlpriblizno}, relation \eref{eq:wqasim} is obtained, with coefficients $a_{ij}$  equal to
\begin{eqnarray}
 a_{11}&=& k_B\left(\frac{m_{k_B}}{p_{k_E}}\right)^3\, , \quad     a_{12} =2(k_C+3)\left(\frac{m_{k_B}}{p_{k_E}}\right)^3\, , \nonumber\\ a_{21}&=&\frac{k_E}{2}\left(\frac{m_{k_B}}{p_{k_E}}\right)^3\, ,
\quad a_{22}=(k_C+6)\left(\frac{m_{k_B}}{p_{k_E}}\right)^3\, , \nonumber
\end{eqnarray}
and
\[b_{13}=\frac{s_{k_C+4}^{CC}}{p_{k_E}}\left(\frac{m_{k_B}}{p_{k_E}}\right)^2\, .\]

\section{Maximally isolated sites within maximally entangled HWs \label{app:diskusija}}

In this Appendix we want to derive the formulae \eref{eq:nli} and \eref{eq:nlio} for the numbers $N_l^{CI}$ and $N_l^{OI}$ of maximally isolated sites within the maximally entangled closed (MEC) and maximally entangled open (MEO) Hamiltonian walks, respectively. Since  MEC HWs on the generator $G_l(b)$ of order $l$ are closed HWs with the maximal number of $E$-steps, and maximally isolated sites within MEC HW are "interior" vertices of $B$-steps made through unit tetrahedrons ({\em i.e.} not the vertices at which HW enters or exits the tetrahedron), number $N_l^{CI}$ is equal to
\begin{equation}
N_l^{CI}=2N_l^B\, ,\end{equation}
where $N_l^B$ is the number of unit tetrahedrons traversed by $B$-steps.  In order to calculate $N_l^B$, and consequently $N_l^{CI}$, we observe maximally entangled $B$-type (MEB) HWs, {\em i.e.} $B$-type HWs with the maximal number of $E$-steps, and, similarly, maximally entangled $E$-type (MEE) HWs. For MEB HW passing through $G_{l+1}(b)$ generator, we introduce label $N_{l+1}^{BB}$ for the numbers of unit tetrahedrons traversed by $B$-step. In a similar way, let $N_{l+1}^{EB}$ be the number of unit tetrahedrons of $G_{l+1}(b)$, traversed by $B$-step, for MEE HW. Then, from the recursion relation for the numbers $B_l$ and $E_l$  \eref{eq:eb}, follows the matrix recursion relation for the numbers $N_l^{BB}$ and $N_l^{EB}$:
 \begin{equation}
\left(
  \begin{array}{c}
    N_{l+1}^{BB} \\
    N_{l+1}^{EB} \\
  \end{array}
\right)=\left(
          \begin{array}{cc}
            k_B & N_G-k_B \\
            k_E & N_G-k_{E} \\
          \end{array}
        \right)\left(
                 \begin{array}{c}
                   N_l^{BB} \\
                   N_l^{EB} \\
                 \end{array}
               \right)\, .\end{equation}
 Since $k_B$ and $k_E$ are given by \eref{eq:kbke}, and for the unit tetrahedron one has $N_0^{BB}=1$ and $N_0^{EB}=0$, solving the obtained recursion relation one obtains
 \begin{eqnarray}
 N_l^{BB}&=&\frac{3+k_C}{N_G-3}3^{l}+\frac{N_G-k_C-6}{N_G-3}N_G^{l}\, , \\
 N_l^{EB}&=&-\frac{N_G-k_C-6}{N_G-3}3^{l}+ \frac{N_G-k_C-6}{N_G-3}N_G^{l}\, . \label{eq:nlbb}
 \end{eqnarray}

Taking into account that MEC HW on $G_{l+1}(b)$ consists of $k_C$ MEE HWs on $G_l(b)$ generators and $(N_G-k_C)$ MEB HWs on $G_l(b)$, as implied by \eref{eq:zatvorene}, it follows
\[
N_{l+1}^B=(N_G-k_C)N_l^{BB}+k_CN_l^{EB}\, ,
\]
and consequently
\begin{equation}
N_l^B=\frac{N_G+k_C}{N_G-3}3^{l}+\frac{N_G-k_C-6}{N_G-3}N_G^{l}\, .\end{equation}
Since $N_l^{CI}=2N_l^B$, $N_l=4N_G^l$, and $\sigma=\frac{\ln 3}{\ln N_G}$, as given by  \eref{eq:sigma}, formula \eref{eq:nli} straightforwardly follows. In a similar manner, from \eref{eq:otvorene}, \eref{eq:ac}, \eref{eq:r}, \eref{eq:fxy}, \eref{eq:rgranice}, and \eref{eq:nlbb} for MEO one obtains \eref{eq:nlio}.


\end{document}